\documentclass[11pt,a4paper]{article}
\pdfoutput=1
\usepackage{amsmath,amssymb,dsfont}
\usepackage{graphicx,color}
\usepackage{verbatim, cancel}
\usepackage[bf]{caption}
\usepackage{ulem,stackrel}
\usepackage{epsfig}

\usepackage{epsfig}

\usepackage[breaklinks,bookmarks=true,colorlinks=true,linkcolor=blue,citecolor=blue,urlcolor=blue,hyperfigures=true]{hyperref}
\usepackage[nosort]{cite} %1
\usepackage[bulletsep]{collref} % punctsep

\numberwithin{equation}{section}

\ifx\href\asklfhas\newcommand{\href}[2]{#2}\fi

\definecolor{pink}{rgb}{0.7,0,0.7}
\definecolor{green}{rgb}{0,0.5,0}
\definecolor{orange}{rgb}{1,0.4,0.3}
\newcommand{\be}{\begin{equation}}
\newcommand{\ee}{\end{equation}}
\newcommand{\ba}{\begin{aligned}}
\newcommand{\ea}{\end{aligned}}
\newcommand{\ben}{\begin{displaymath}}
\newcommand{\een}{\end{displaymath}}
\newcommand{\bea}{\begin{eqnarray}}
\newcommand{\eea}{\end{eqnarray}}
\newcommand{\bean}{\begin{eqnarray*}}
\newcommand{\eean}{\end{eqnarray*}}
\newcommand{\bpmat}{\begin{pmatrix}}
\newcommand{\epmat}{\end{pmatrix}}

\newcommand{\AdS}{{\ensuremath{\mathrm{AdS}}}}
\newcommand{\Sph}{{\ensuremath{\mathrm{S}}}}

\newcommand{\PSU}{{\ensuremath{\mathrm{PSU}}}}
\newcommand{\SO}{{\ensuremath{\mathrm{SO}}}}
\newcommand{\SLS}{{\ensuremath{\mathrm{SL}}}}
\newcommand{\SU}{{\ensuremath{\mathrm{SU}}}}
\newcommand{\UN}{{\ensuremath{\mathrm{U}}}}
\newcommand{\OSP}{{\ensuremath{\mathrm{OSP}}}}

\newcommand{\alg}[1]{\ensuremath{\mathfrak{#1}}}

\newcommand{\psu}{{\ensuremath{\mathfrak{psu}}}}
\newcommand{\so}{{\ensuremath{\mathfrak{so}}}}
\newcommand{\sls}{{\ensuremath{\mathfrak{sl}}}}
\newcommand{\su}{{\ensuremath{\mathfrak{su}}}}
\newcommand{\osp}{{\ensuremath{\mathfrak{osp}}}}

\newcommand{\Integers}{{\ensuremath{\mathbb{Z}}}}
\newcommand{\Reals}{{\ensuremath{\mathbb{R}}}}

\newcommand{\calN}{{\ensuremath{\mathcal{N}}}}
\newcommand{\calA}{{\ensuremath{\mathcal{A}}}}
\newcommand{\calB}{{\ensuremath{\mathcal{B}}}}

\renewcommand{\l}{\lambda}
\renewcommand{\th}{\theta}
\renewcommand{\a}{\alpha}
\renewcommand{\b}{\beta}
\newcommand{\g}{\gamma}

\renewcommand{\d}{\delta}
\newcommand{\eps}{\epsilon}
\newcommand{\veps}{\varepsilon}

\newcommand{\Th}{\Theta}
\newcommand{\m}{\mu}
\newcommand{\n}{\nu}
\renewcommand{\k}{\kappa}
\newcommand{\vk}{\varkappa}

\newcommand{\f}{\frac}
\newcommand{\p}{\partial}

\newcommand{\diag}{\ensuremath{\mathrm{diag}}}
\newcommand{\Span}{\ensuremath{\mathrm{span}}}

\newcommand{\la}{\langle}
\newcommand{\La}{\left\langle}
\newcommand{\ra}{\rangle}
\newcommand{\Ra}{\right\rangle}

\renewcommand{\tt}{\ensuremath{\mathbf{t}}}%overwrites \tt, similar to \texttt
\newcommand{\TT}{\ensuremath{\mathbf{T}}}

\newcommand{\dd}{{\ensuremath{\text{d}}}}

\long\def\symbolfootnote[#1]#2{\begingroup
\def\thefootnote{\fnsymbol{footnote}}\footnote[#1]{#2}\endgroup}

\begin{document}

\begin{titlepage}
%\vspace{10pt}
\hfill\parbox[t]{4cm}{
\texttt{FU-Ph 10/2016~(06) \\
NORDITA-2016-100 \\
ZMP-HH/16-22}
}

\vspace{20mm}

\begin{center}

{\Large\bf Quantization of the ${\rm AdS}_3$ Superparticle on ${\rm OSP}(1|2)^2/{\rm SL}(2,\mathbb{R})~$}

\vspace{30pt}

{Martin Heinze,$^{a,\,b}$ George Jorjadze,$^{c,\,d}~$ 
%Luka Megrelidze,$^d$
%\symbolfootnote[2]{\tt{\{jorj,jan.plefka,pollok\}@physik.hu-berlin.de}}
}
\\[6mm]

{\small
{\it\ ${}^a$II. Institut f{\"u}r Theoretische Physik, Universit{\"a}t Hamburg,\\
	Luruper Chaussee 149, 22671 Hamburg, Germany}\\[2mm]
{\it\ ${}^b$Zentrum f\"ur Mathematische Physik, Universit\"at Hamburg,\\ 
	Bundesstrasse 55, 20146 Hamburg, Germany}\\[2mm]
{\it${}^c$Free University of Tbilisi,\\
		Agmashenebeli Alley 240, 0159, Tbilisi, Georgia}\\[2mm]
{\it${}^d$Razmadze Mathematical Institute of TSU,\\
Tamarashvili 6, 0177, Tbilisi, Georgia}\\[5mm]
\texttt{martin.heinze@desy.de},\qquad\texttt{george.jorjadze@physik.hu-berlin.de},
%[2mm]
% {\it${}^d$Ilia State University,\\
%K. Cholokashvili Ave 3/5, 0162, Tbilisi, Georgia}\\
}

\vspace{30pt}

\end{center}

\centerline{{\bf{Abstract}}}
\vspace*{5mm}
\noindent
	We analyze ${\rm AdS}_3$ superparticle dynamics on the coset ${\rm OSP}(1|2) \times {\rm OSP}(1|2)/{\rm SL}(2,\Reals)$.  The system is quantized in canonical coordinates obtained by gauge invariant Hamiltonian reduction. 
	The left and right Noether charges of a massive particle are parametrized by coadjoint orbits of a timelike element of $\frak{osp}(1|2)$. Each chiral sector is described by two bosonic and two fermionic canonical coordinates corresponding to a superparticle with superpotential $W=q-m/q$, where $m$ is the particle mass. 
	Canonical quantization then provides a quantum realization of $\frak{osp}(1|2)\oplus\frak{osp}(1|2)$.
	For the massless particle the chiral charges lie on the coadjoint orbit of a nilpotent element of $\frak{osp}(1|2)$ and each of them depends only on one real fermion, which demonstrates the underlying $\k$-symmetry.
	These remaining left and right fermionic variables form a canonical pair and the system is described by four bosonic and two fermionic canonical coordinates. Due to conformal invariance of the massless particle, the $\frak{osp}(1|2)\oplus\frak{osp} (1|2)$ extends to the corresponding superconformal algebra  $\frak{osp}(2|4)$. Its 19 charges are given by all real quadratic combinations of the canonical coordinates, which trivializes their quantization.

\vspace{4cm}

\end{titlepage}

{\hypersetup{linkcolor=black}
\tableofcontents}

\section{Introduction}\label{sec:Intro}
	For more than a decade the existence of integrability in the AdS/CFT correspondence has excited astonishing insights into non-perturbative aspects of both conformal field theories (CFT) as well as string theories in Anti-de Sitter space ($\AdS$) \cite{Arutyunov:2009ga, Beisert:2010jr, Bombardelli:2016rwb}. In particular, unraveled first for the duality between $\calN = 4$ super Yang-Mills theory and the $\AdS_5\times\Sph^5$ superstring, the conjectured quantum integrability has allowed for a solution of the spectral problem through the mirror Thermodynamic Bethe Ansatz (TBA) \cite{Arutyunov:2009ur, Bombardelli:2009ns, Gromov:2009tv, Gromov:2009bc} as well as the Quantum Spectral Curve \cite{Gromov:2013pga}, which ostensibly amounts to quantization of the system. 

	However, in spite of this progress it is worth noting that our understanding of quantization of the $\AdS_5\times\Sph^5$ superstrings from first principles is still limited. The spectrum of the $1/2$-BPS subsector, {\it viz.} of the corresponding supergravity, is well-known \cite{Kim:1985ez, Gunaydin:1984fk} and, using the results of \cite{Metsaev:1999gz, Metsaev:2000yu}, it was shown to match with quantization of the massless $\AdS_5\times\Sph^5$ superparticle \cite{Horigane:2009qb}, see also \cite{Siegel:2010gm} as well as the recent work on the supertwistor formulation \cite{Arvanitakis:2016vnp}.
	In fact, it seems favorable to attain a rigorous understanding of the massless superparticle before attempting to quantize the superstring. 

	Another well-studied sector is the class of {\it heavy}, respectively, {\it long} string states captured by semi-classical string solutions. As in the seminal works \cite{Berenstein:2002jq, Gubser:2002tv, Frolov:2002av, Frolov:2003qc,  Arutyunov:2003uj}, here one relies on some of the global $\psu(2,2|4)$ charges to diverge in the 't Hooft coupling as $\sqrt{\l}$, resulting in a similar scaling of the string energy, $E \propto \sqrt{\l}$.
	Fluctuations around such configurations can then be quantized perturbatively. For instance, fluctuations around the point particle of diverging $\Sph^5$ momentum are described by the BMN string \cite{Berenstein:2002jq} and the corresponding quantum corrections were calculated in \cite{Callan:2003xr, Callan:2004uv, Callan:2004ev, Frolov:2006cc}, which allowed to construct the scattering $S$-matrix in this limit \cite{Arutyunov:2006yd, Klose:2006zd, Klose:2007rz}. 

	For {\it light}, respectively, {\it short} string states with finite $\psu(2,2|4)$ charges, however, such a perturbative description formally breaks down and it has been a renowned problem to obtain the spectrum beyond the leading order \cite{Gubser:2002tv}, $E \propto {\l}^{1/4}$. The difficulties seem to be caused by the particular scaling of the string zero-modes \cite{Passerini:2010xc}, {\it viz.} the particle-like degrees of freedom of the center-of-mass.
	At the same time, this points out that for short strings the customary uniform light cone gauge \cite{Arutyunov:2004yx, Arutyunov:2005hd} might not be the most appropriate gauge choice.

	Therefore, restricting to bosonic $\AdS_5\times\Sph^5$ and employed static gauge \cite{Jorjadze:2012iy}, in \cite{Frolov:2013lva} a semi-classical string solution has been constructed generalizing the pulsating string \cite{deVega:1994yz, Minahan:2002rc} by allowing for unconstrained zero-modes. Apart from showing classical integrability and invariance under the isometries $\SO(2,4)\times \SO(6)$, the energy of the lowest excitation of this so called single-mode string proved to match with integrability based results for the Konishi anomalous dimension up the first quantum corrections, the order $\l^{-1/4}$.
	For this the crucial step has been to reformulate the system as a massive $\AdS_5\times\Sph^5$ particle \cite{Dorn:2005ja, Dorn:2010wt} with the mass term determined by the stringy non-zero-modes. Hence, in order to understand quantization of the $\AdS$ superstrings from first principles it seems favorable to study not only massless but also massive $\AdS$ superparticles.

	Notably, the previous observation is equivalent to the statement that the single-mode string \cite{Frolov:2013lva} is the ${\SO}(2,4)\times {\SO}(6)$ orbit of the pulsating string \cite{deVega:1994yz, Minahan:2002rc}. This suggests to construct (super)isometry group orbits of other semi-classical string solutions, which has the additional appeal that the Kirillov-Kostant-Souriau method of coadjoint orbits yields a quantization scheme in terms of the symmetry generators, which is manifestly gauge-independent.
	In \cite{Heinze:2014cga} we followed this idea by constructing the isometry group orbits of the bosonic particle and spinning string in $\AdS_3\times\Sph^3$, leading to a Holstein-Primakoff realization for the isometry algebra \cite{Holstein:1940zp, Dzhordzhadze:1994np, Jorjadze:2012jk} in agreement with previous results. We then turned to superisometry group orbits by applying orbit method quantization to the $\AdS_2$ superparticle on $\OSP(1|2)/\SO(1,1)$ \cite{Heinze:2015oha}, yielding a Holstein-Primakoff-like realization of the superisometries $\osp(1|2)$. For the massless case however the $\k$-symmetry transformation left only one physical real fermion, rendering the model quantum inconsistent.

	In this work we continue this program and apply superisometry group orbit quantization to the $\calN = 1$ superparticle on the $\AdS_3$ superspace defined on the coset\footnote{Here and in the following, we abbreviate the direct product of supergroups $G$ as $G \times G = G^2$.} $\OSP(1|2)^2/\SLS(2,\Reals)$. More specifically, we will investigate the action showing $\k$-symmetry in the massless case, as it constitutes a truncation of the Green-Schwarz superstring encountered in the AdS/CFT correspondence. Additionally, we will demonstrate that only for this $\k$-symmetric action there is a close relation to the superparticle on the supergroup $\OSP(1|2)$, a statement which carries over to general cosets of the form $G^2/H$. 

	Let us note already that in comparison to \cite{Heinze:2015oha} the present coset exactly doubles the number of fermionic degrees of freedom. Hence, by construction we are circumventing the problems encountered in the massless case of \cite{Heinze:2015oha}, as now $\k$-symmetry will leave us with two real fermions, enough to form one fermionic canonical pair.
	Therefore, this model amounts to what is arguably the simplest quantum consistent massless $\AdS$ superparticle.\footnote{Contestants to this title might be the $\AdS_2$ superparticle actions on $\SU(1,1|1)/(\SO(1,1)\times\UN(1))$ or $\SU(1,1|1)/\SO(1,1)$, see for example \cite{Bellucci:2002va, Ivanov:2002tb} and the more recent works \cite{Galajinsky:2010zy, Galajinsky:2011xp, Bellucci:2011hk, Orekhov:2014xra}, as well as a non-$\k$-symmetric version of the $\AdS_2$ superparticles on $\OSP(1|2)/\SO(1,1)$.} 

	Indeed, for both the massive and the massless superparticle, by using the orbit method we will obtain not only physical canonical variables, which can be quantized in terms of bosonic and fermionic oscillators, but also conserved charges forming a Holstein-Primakoff-like quantum realization \cite{Heinze:2015oha} of the superisometry algebra $\osp_l(1|2)\oplus\osp_r(1|2)$.

	For the massive case we point out that both chiral subsectors are described by supersymmetric quantum mechanics with superpotential $W=q-\f{2\m-1/2}{q}\,$ \cite{Cooper:1994eh}.
	
	For massless particles it is well-known that the action is invariant not only under the isometries but under the full conformal symmetries of the underlying space-time. For $\AdS_{N+1}$ this yields an extension of the isometry algebra $\so(2,N)$ to the conformal algebra $\so(2,N+1)$ \cite{Dorn:2005vh}.
	Correspondingly, for the massless superparticle at hand we find that the superisometries $\osp_l(1|2)\oplus\osp_r(1|2)$ extend to the superconformal algebra $\osp(2|4)$.

	This work has clearly been motivated by and is aimed towards a future application to semi-classical string solutions of the $\AdS_5 \times \Sph^5$ superstring. However, there is actually a whole plethora of semi-symmetric $\AdS$ supercoset \cite{Zarembo:2010sg} which might serve as backgrounds for integrable sigma-models encountered in the AdS/CFT correspondence. 
	In particular, initiated by \cite{Babichenko:2009dk, David:2008yk, OhlssonSax:2011ms, Sundin:2012gc, Cagnazzo:2012se, Borsato:2012ud, Borsato:2013qpa} there has been remarkable  progress on the $\AdS_3$/CFT${}_2$ correspondence on $\AdS_3\times\Sph^3\times\mathrm{T}^4$ and $\AdS_3\times\Sph^3\times\Sph^3\times\Sph^1$, see also the more recent works \cite{Abbott:2015mla, Abbott:2015pps, Sundin:2016gqe, Stromwall:2016dyw, Borsato:2016xns, Fontanella:2016opq} as well as the review \cite{Sfondrini:2014via}.\footnote{Further studies on $\AdS_3$ superstrings, especially in the RNS description, include \cite{Giveon:1998ns, Elitzur:1998mm, deBoer:1999gea, Gukov:2004ym}, see also the more recent works \cite{Gaberdiel:2013vva, Tong:2014yna}. } The $\AdS_3$ superparticle under investigation is naturally viewed as a truncation of these string theories.
	Similarly, this work might also prove relevant for supersymmetric versions of the non-critical $\AdS_3$ string \cite{Brown:1986nw}, see also \cite{Maldacena:2000hw} as well as the work \cite{Hikida:2007sz} on the $\OSP(1|2)$ WZNW model, and even of the $\AdS_3$ higher spin theory \cite{Vasiliev:1995dn, Vasiliev:1999ba}.

	Particle dynamics in $\AdS_3$ (super)space have also been investigated in a series of other works. The dynamical sectors of the bosonic $\AdS_3$ particle were investigated in \cite{Batlle:2014sca}, where for critical spin $J = J_{1 2} = m$ the systems reduces to a particle on $\AdS_2$. Higher derivative actions for the $\AdS_3$ superparticle on $\SU(1,1|1)$ were derived in \cite{Kozyrev:2016mlo}, see also \cite{Kozyrev:2013vla}, and similar techniques have been applied to multi-particle  dynamics, see \cite{Krivonos:2010zy, Galajinsky:2016wuc} and references therein, which are relevant for the duality between black holes and superconformal Calogero models \cite{Claus:1998ts, Gibbons:1998fa}.

	The paper is organized as follows. In Section \ref{sec:Bos} we study the bosonic $\AdS_3$ particle on $\SLS(2,\Reals)$. After establishing the isometry between $\AdS_3$ and $\SLS(2,\Reals)$ and the $\AdS_3$ conformal algebra we discuss the massive and massless particle dynamics. In Section \ref{sec:Super} we then turn to the $\AdS_3$ superparticle on $\OSP(1|2)^2/\SLS(2,\Reals)$. Here, we first discuss the coset construction to then study the massive and massless case. A conclusion and outlook are given in Section \ref{sec:concl}. Some technical details of the calculations are collected in three appendices.

\section{The bosonic \texorpdfstring{$\AdS_3$}{AdS3} particle}% on \texorpdfstring{$\SLS(2,\Reals)$}{SL2R} }
\label{sec:Bos}

\subsection{Isometry between \texorpdfstring{$\AdS_3$}{AdS3} and \texorpdfstring{$\SLS(2,\Reals)$}{SL2R}}\label{subsec:BosIsom}

	Let us consider the space $\Reals^{2,2}$ with coordinates $X^A,$ $\,A=0', 0, 1, 2\,$, and the metric tensor $\eta_{AB}=\diag(-1,-1,1,1)$. The hyperboloid embedded in $\Reals^{2,2}$,
	\be\label{hyperbola}
		X^A X_A +1=0~,
	\ee
	is identified with $\AdS_3$
	and its map to the $\SLS(2,\Reals)$ group manifold is given by
	\be\label{g=X}
		g=\begin{pmatrix}
			X^{0'}+X^2 &X^1+X^0\\X^1-X^0& X^{0'}-X^2
			\end{pmatrix}\,. 
	\ee
	This group element and its inverse can be written as 
	\be\label{g=X1}
		g= X^{0'}\,{\bf I}+X^\mu\,\tt_\mu~, \qquad\qquad
			g^{-1}= X^{0'}\,{\bf I}-X^\mu\,\tt_\mu~,
	\ee
	where ${\bf I}$ is the unit matrix and
	$\tt_\m$ for $\,\m=0,1,2\,$ form a basis of $\sls(2,\Reals)$,
	\be \label{sl(2) basis}
		{\bf{t}}_0=\begin{pmatrix} 0&1\\-1&0 \end{pmatrix}\,,\qquad
		{\bf{t}}_1=\begin{pmatrix} \,0&1\\ \,1&0 \end{pmatrix}\,,\qquad
		{\bf{t}}_2=\begin{pmatrix} \,1&0\\ \,0&-1 \end{pmatrix}\,.
	\ee

	The commutation relations of the basis vectors is given by
	\be\label{tt comm}
		[{\bf{t}}_\mu ,\,{\bf{t}}_\nu]=2\eps_{\mu\nu}\,^\rho\,{\bf{t}}_\rho~
	\ee
	where $\eps_{\mu\nu\rho}$ is the Levi-Civita tensor,
	$\eps_{012}=1$. Here, rising and lowering of indices is provided by the metric tensor $\eta_{\mu\nu}=\mbox{diag}(-1,1,1)$,
	which corresponds to the inner product in $\sls(2,\Reals)$ defined by
	$\langle\, \tt_\mu\,\tt_\nu\,\rangle \equiv \frac{1}{2}\,
	\mbox{tr}(\tt_\mu\,\tt_\nu)=\eta_{\mu\nu}$. With the help of \eqref{g=X1}  and \eqref{hyperbola} one then obtains  
	the isometry between $\SLS(2,\Reals)$ and $\AdS_3$,
	\be\label{dg=dX}
		\langle(\dd g\,g^{-1})\,(\dd g\,g^{-1})\rangle = \dd X^A \dd X_A~.
	\ee
	
	The $\SO(2,2)$ isometry group of $\AdS_3$ is generated by the infinitesimal transformations
	\be\label{AdS isometry tr}
		X^A\mapsto X^A+\a^A{}_B\,X^B~,\qquad\qquad \alpha^{AB}=-\alpha^{BA}~.
	\ee
	In $\SLS(2,\Reals)$, these correspond to left-right multiplication of the group element 
	\be\label{SL(2) isometry tr}
		g\mapsto g+\a^{\n}_l\,\tt_\n\,g+\a^{\n}_r\,g\,\tt_\n~,
	\ee
	with $\a^\n_l=\f 1 2 \a^{0'\n}+\f 1 4 \eps^\n\,_{\rho\m}\,\a^{\rho\m}$ and 
	$\a^\n_r=\f 1 2\a^{0'\n}-\f 1 4 \eps^\n\,_{\rho\m}\,\a^{\rho\m}$ ({\it cf.} Appendix \ref{app:SL2R}), hence establishing the algebra isomorphism $\so(2,2) = \sls_l(2,\Reals) \oplus \sls_r(2,\Reals)$.

	The infinitesimal conformal transformation of $\AdS_3$ read \cite{Dorn:2005vh}
	\be\label{AdS conformal tr}
		X^A\mapsto X^A+\veps^B(\d_B{}^A+X_B X^A)~,
	\ee
	leading to a rescaling of the metric \eqref{dg=dX} by the factor 
	$\,1+2\,\veps^B\, X_B\,$. Correspondingly, these extend the isometry algebra $\so(2,2)$ to the $\AdS_3$ conformal algebra $\so(2,3)$. For the $\SLS(2,\Reals)$ group element \eqref{g=X1} the transformations \eqref{AdS conformal tr} take the form
	(see also \eqref{SL(2) conformal tr 1})
	\be\label{SL(2) conformal tr}
		g\mapsto g+\veps^{0'}{\bf I} +\veps^{\m}\tt_\m+(\veps^A X_A) g~.
	\ee

\subsection{Particle dynamics on \texorpdfstring{$\SLS(2,\Reals)$}{SL2R}}

	The dynamics  of a particle in $\SLS(2,\Reals)$ can be described by the action
	\be\label{action 1}
		S=\int \dd \tau\,\left(\frac{1}{2\,\xi} \La\,\dot{g}\,g^{-1}\,\dot{g}\,g^{-1}\,\Ra
		-\frac{\xi\,m^2}{2}\right)~.
	\ee
	Here, $\tau$ is an evolution parameter, $\xi$ plays the role of a worldline einbein 
	and $m$ is a particle mass.  The isometry transformations \eqref{SL(2) isometry tr} 
	yield the Noether charges
	\be\label{R-L}
		L=\f{\dot g\,g^{-1}}{\xi}~, \qquad\qquad R=\f{g^{-1}\dot g}{\xi}~,
	\ee
	which are related by $L=g R g^{-1}$ and therefore 
	have the same length, $\la\, L\,L\, \ra=\la\, R\,R\, \ra$.

	In the first order formalism the action \eqref{action 1} is equivalent to
	\be\label{action 2}
		S=\int \dd \tau\,\left(\La L\,\dot{g}\,g^{-1}\Ra
		- \frac{\xi}{2} \left(\La L\,L\Ra + m^2 \right)\right)~,
	\ee
	which leads to the Hamilton equations 
	\be\label{Hamilton eq}
		\dot g g^{-1}\,=\xi L~,   \qquad\qquad \dot L=0~,   
	\ee
	and the mass-shell condition
	\be\label{mass-shell}
		\la L\,L \ra +m^2 =0~.
	\ee

	We use the Faddeev-Jackiw method that reduces the system to the physical degrees of freedom. 
	The reduction schemes for the massive and the massless cases are different.

\subsection{Massive particle on \texorpdfstring{$\SLS(2,\Reals)$}{SL2R}}
	We first analyze the  massive case, which
	corresponds to timelike $L$ and $R$. Due to the mass-shell condition \eqref{mass-shell} they are on the adjoint orbit of the $\sls(2,\Reals)$ element $m\tt_0$ and one can use the parametrization
	\be\label{L=t0,R=t0}
		L=m\, g_l\,\tt_0\,g_l^{-1}~, \qquad R=m\, g_r^{-1}\,\tt_0\,g_r~, \qquad g=g_l\,g_r~.
	\ee
	The presymplectic form $\Th=\la\,L\, \dd g\,g^{-1}\,\ra$ then splits 
	into the sum of left and right parts
	\be\label{1-form m}
		\Th=m\la\,\tt_0\,g_l^{-1}\,\dd g_l\,\ra+m\la\,\tt_0\,\dd g_r\,g_r^{-1}\,\ra~.
	\ee
	Defining the nilpotent generators $\tt_\pm=\f{1}{2}(\tt_1\pm\tt_0)$, we use the Iwasawa decomposition
	\be\label{Iwasawa}
		g_l=e^{\g_l\,\tt_+}\,e^{\a_l\,\tt_2}\,e^{\th_l\,\tt_0}~, \qquad\qquad 
		g_r=e^{\th_r\,\tt_0}\,e^{\a_r\,\tt_2}\,e^{\g_r\,\tt_+}~,
	\ee
	see also Appendix \ref{app:SL2R}.
	Plugging this parametrization into \eqref{1-form m}, we find the presymplectic form
	\be\label{SF dTh}
		\Th=-m\,\dd \th_l-\f m 2\,e^{-2\a_l}\dd \g_l-m\,\dd \th_r-\f m 2\,e^{2\a_r}\dd \g_r~,
	\ee
	and the Noether charges \eqref{L=t0,R=t0}
	\be\label{L,R in angles}
		L=m\begin{pmatrix}
				-\g_l\,e^{-2\a_l} & \g_l^2\,e^{-2\a_l}+e^{2\a_l} \\ -e^{-2\a_l}& \g_l\,e^{-2\a_l}
			\end{pmatrix}\,,\qquad 
		R=m\begin{pmatrix}
				\g_r\,e^{2\a_r} & \g_r^2\,e^{2\a_r}+e^{-2\a_r} \\ -e^{2\a_r}& -\g_r\,e^{2\a_r}
			\end{pmatrix}\,,
	\ee
	hence rendering $\th_l$ and $\th_r$ unphysical. Introducing the canonical coordinates by 
	\be\label{CC m}
		p_l=\sqrt m\,\g_l\,e^{-\a_l}~,\quad q_l=\sqrt m\,e^{-\a_l}, \qquad  
		p_r=\sqrt m\,\g_r\,e^{\a_r}~,\quad q_r=\sqrt m\,e^{\a_r},
	\ee
	from \eqref{SF dTh} and \eqref{L,R in angles} we find 
	$\dd \Th=\dd p_l\wedge \dd q_l+\dd p_r\wedge \dd q_r$ 
	and  
	\be\label{L-R in pq m}
		L=\begin{pmatrix}
			-p_l q_l & p_l^2+m^2\,q^{-2}_l \\ -q_l^2& p_lq_l
			\end{pmatrix}\,,\qquad \qquad
		R=\begin{pmatrix}
			p_r q_r & p_r^2 +m^2\,q^{-2}_r\\ -q_r^2 & -p_r q_r
		\end{pmatrix}\,.\quad
	\ee
	The dynamical integrals 
	$L_\m=\la\,\tt_\m\, L\,\ra\,$ and $\,R_\m=\la\,\tt_\m\, R\,\ra$ then take the form 
	\be\ba\label{DI m>0}
		&L^0=\f 1 2(p_l^2+q_l^2)+\f{m^2}{2q_l^2}~,& \qquad &L_1=\f 1 2(p_l^2-q_l^2)+\f{m^2}{2q_l^2}~,& \qquad &L_2=-p_lq_l~,&\\
		&R^0=\f 1 2(p_r^2+q_r^2)+\f{m^2}{2q_r^2}~,& &R_1=\f 1 2(p_r^2-q_r^2)+\f{m^2}{2q_r^2}~,& &R_2= p_rq_r~,&
	\ea\ee
	and their Poisson brackets form the $\sls(2,\Reals)$ algebra \eqref{tt comm}   
	\be\label{PB L-R}
		\{L_\m, L_\n\}=-2\eps_{\m\n}\,^\rho\,L_\rho~, \qquad 
		\{R_\m, R_\n\}=2\eps_{\m\n}\,^\rho\,R_\rho~,
	\ee
	which reflects the isometry invariance on the mass-shell.

	The time translation parameter in \eqref{AdS isometry tr} is $\a^{0'}\,_0$ and due to
	\eqref{SL(2) isometry tr} the energy reads 
	\be\label{Energy}
		E=\f 1 2(L^0+R^0) ~.%= \frac{1}{4}\Big(\vec{p}^{\,2} + \vec{q}^{\,2} + \frac{m^2}{\vec{q}^{\,2}}\Big)~,
	\ee
	%with $\vec{p}=(p_l,p_r)$ and $\vec{q}=(q_l,q_r)$ and {\it e.g.} $\vec{q}^{\,2} = q_l^2 +q_r^2$.
	%Note that $E$ resembles the Hamiltonian of the Neumann-Rosochatius (NR) system, see also \cite{Arutyunov:2003za}. However, $\vec{q}$ does not necessarily lie on an $\Sph^1$, {\it i.e.} we do not impose the NR constraints $\vec{q}^{\,2} =1$ and $\vec{q}\cdot\vec{p}=0$.

	Now we describe quantization of the system \eqref{DI m>0}-\eqref{PB L-R}. Since the canonical coordinates \eqref{CC m} are given on the half-planes ($q_l>0$, $q_r>0$),
	it is natural to quantize the system in the coordinate representation.
	Thus, only the charges $L_2$ and $R_2$ exhibit ambiguous operator ordering. 
	A quantum realization of the algebra \eqref{PB L-R} is provided by the Weyl ordering and the energy spectrum is obtained from the analysis of the eigenvalue problem for the operator 
	\be\label{Hamiltonian}
		H=\f 1 4(-\partial^2_{q}+q^2 +m^2/q^2)~.
	\ee

	Due to the commutation relations of $\sls(2,\Reals)$, the operators 
	\be\label{J_pm}
		J^\pm=\f 1 4(-\partial^2_{q}-q^2+m^2/q^2)\pm \f 1 2(q\p_q+1/2)
	\ee
	are rising and lowering for $H$, {\it i.e.} $[H,J^\pm]=\pm J^\pm$. $\,H$ then has the harmonic oscillator spectrum   
	with some minimal eigenvalue $\m$ and
	the ground state wave function $\Psi_0(q)$ satisfies the equations $H\Psi_0=\m\,\Psi_0$ and $J^-\Psi_0=0$.

	Derivation of the eigenfunctions is simplified due to the relation
	\be\label{J_pm1}
		J^\pm= H-\f{q^2}{2}\pm\f1 2(q\p_q+1/2)~,
	\ee
	which leads to a first order differential equation for $\Psi_0(q)$. 
	Up to a normalization constant, one simply obtains
	\be\label{psi0}
		\Psi_0(q)\,\propto\, q^{2\m-\f  1 2}\,\, e^{-\f{1}{2}\,q^2}
	\ee
	and the minimal eigenvalue $\m$ is related to the mass parameter by\,\footnote{Two possible signs
	in \eqref{E min} correspond to two different self-adjoint extensions of the operator \eqref{Hamiltonian}
	valid for $m^2\in [-1/4, <3/4)$  \cite{Jorjadze:2012jk}, see also the massless case below.}
	\be\label{E min}
	\m=\f{1\pm\sqrt{m^2+1/4}}{2}~.
	\ee
	Note that the ground state wave function \eqref{psi0} is normalizable for  $\m>0$, which
	corresponds to the unitarity bound in $\AdS_3$ \cite{Breitenlohner:1982jf}.

	Higher level eigenstates are obtained by acting with the rising operator $J^+$ \eqref{J_pm1}, yielding
	\be\label{Psi_n}
		\Psi_n\, \propto\, P_n(q^2)\, q^{2\m-\f 1 2}\, e^{-\f{1}{2}\,q^2}~,
	\ee
	with the recursive relations $P_{n+1}(x)=(2\m+n-x)P_n(x)+xP'_n(x)$, such that after a suitable normalization $P_n(x)$ become generalized Laguerre polynomials.

	The left and right copies of the generators \eqref{Hamiltonian}-\eqref{J_pm} form a representation of $\sls(2,\Reals)\oplus \sls(2,\Reals)$, which is unitary equivalent to the Holstein-Primakoff type representation obtained in \cite{Dzhordzhadze:1994np}, see also \cite{Heinze:2014cga, Jorjadze:2015}.
	The Holstein-Primakoff representation of $\sls(2,\Reals)$ reads
	\be\label{Holstein-Pr}
		E=\m+b^\dag b~,  \quad B^\dag=b^\dag\,\sqrt{2\m+b^\dag b}~, \quad B=\sqrt{2\m+b^\dag b}\,\,b~,
	\ee
	where $b^\dag$ and  $b$ are the standard creation-annihilation operators with canonical commutation $[b,b^\dag]=1$ and to recover \eqref{Hamiltonian}-\eqref{J_pm} we employ the unitary map $E\mapsto H$, $B^\dag\mapsto -J^+$ and $B\mapsto -J^- $. 
	The corresponding canonical transformation
	can be found in Appendix \ref{app:HolPrim}.

	Furthermore, note that the operators \eqref{Hamiltonian}-\eqref{J_pm} can be written as
	\be\label{H-J} 
	H=\m+A^+ A^-~,  \qquad J^+=A^+(A^- -q)~, \qquad J^-= (A^+ -q)A^-~,
	\ee
	where
	\be\label{A op} 
	A^+ = \f 1 2\left(q-\f{2\m-\f1 2}{q}-\p_q\right)~, \qquad
	A^-=\f 1 2\left(q-\f{2\m-\f1 2}{q} +\p_q\right)~.
	\ee
	This form of the Hamiltonian prepares the system for a supersymmetric extension \cite{Cooper:1994eh}, with superpotential $W=q-\f{2\m-1/2}{q}\,$.

\subsection{Massless particle on \texorpdfstring{$\SLS(2,\Reals)$}{SL2R}}

	The massless case corresponds to
	lightlike $L$ and $R$. These are on the orbit of the nilpotent element, say $\tt_+$, implying the parametrization
	\be\label{L=t+,R=t+}
		L=g_l\,\tt_+\,g_l^{-1}~, \qquad R= g_r^{-1}\,\tt_+\,g_r~, \qquad g=g_l\,g_r~.
	\ee
	Analogously to \eqref{1-form m}, the presymplectic one-form becomes
	\be\label{1-form}
		\Th=\la\,\tt_+\,g_l^{-1}\,\dd g_l\,\ra+\la\,\tt_+\,\dd g_r\,g_r^{-1}\,\ra~.
	\ee
	With the help of the Iwasawa decompositions
	\be\label{gl,gr}
		g_l=e^{\th_l\,\tt_0}\,e^{\a_l\,\tt_2}\,e^{\g_l\,\tt_+}~, \qquad\qquad
		g_r=e^{\g_r\,\tt_+}\,e^{\a_r\,\tt_2}\,e^{\th_r\,\tt_0}~,
	\ee
	we find the one-form (see Appendix \ref{app:SL2R} for details of calculation)
	\be\label{1-form m=0}
	\Th=-\f 1 2\,e^{2\a_l}\dd \th_l-\f 1 2\,e^{-2\a_r}\dd \th_r~,
	\ee 
	and the Noether charges \eqref{L=t+,R=t+}
	\be\ba\label{L,R m=0 in angles}
		L=&\begin{pmatrix}
			e^{2\a_l}\cos\th_l\,\sin\th_l & e^{2\a_l}\cos^2\th_l\\
			-e^{2\a_l}\,\sin^2\th_l & -e^{2\a_l}\cos\th_l\sin\th_l\end{pmatrix},& \\[2mm]
		R=&\begin{pmatrix}
			-e^{-2\a_r}\cos\th_r\sin\th_r & e^{-2\a_r}\cos^2\th_r \\ 
			-e^{-2\a_r}\sin^2\th_r& e^{-2\a_r}\cos\th_r\sin\th_r \end{pmatrix},&
	\ea\ee
	yielding that $\g_l$ and $\g_r$ are unphysical.

	The parameters $\th_l$ and $\th_r$ are cyclic variables and global canonical coordinates read 
	\be\label{CC m=0}
		p_l=e^{\a_l}\,\cos\th_l, \quad  q_l=-e^{\a_l}\,\sin\th_l,   \qquad 
			p_r=-e^{-\a_r}\,\cos\th_r, \quad q_r=e^{-\a_r}\,\sin\th_r.
	\ee
	This provides $\dd \Th=\dd p_l\wedge \dd q_l+\dd p_r\wedge \dd q_r$ 
	and the Noether charges in \eqref{L=t+,R=t+} take the form  
	\be\label{L-R in pq}
		L=\begin{pmatrix} -p_l q_l & p_l^2 \\ -q_l^2& p_lq_l\end{pmatrix},\qquad \qquad
		R=\begin{pmatrix} p_r q_r & p_r^2 \\ -q_r^2 & -p_r q_r\end{pmatrix}~.
	\ee
	Then, similarly to \eqref{DI m>0}, one gets the dynamical integrals
	\be\ba\label{DI m=0}
		&L^0=\f 1 2(p_l^2+q_l^2)~,& \qquad &L_1=\f 1 2(p_l^2-q_l^2)~,& \qquad &L_2=-p_lq_l~,&\\
		&R^0=\f 1 2(p_r^2+q_r^2)~,&       &R_1=\f 1 2(p_r^2-q_r^2)~,&        &R_2= p_rq_r~,&
	\ea\ee
	which form the same Poisson brackets algebra \eqref{PB L-R}.
	Formally, these dynamical integrals are obtained from \eqref{DI m>0} at $m=0$. However, it has to be noticed that the canonical coordinates in \eqref{DI m=0} are given on the full planes without the origin, whereas in the massive case they are defined on the half-planes $(q_l>0,\,q_r>0)$.

	In the massless case there are additional Noether charges $C_A$  related to the conformal 
	transformations \eqref{SL(2) conformal tr}, which yield  $C_{0'}=\la\,g^{-1}L\,\ra,\,$  and
	$\,C_\m=\la\,g^{-1}L\,\tt_\m\,\ra$. These dynamical integrals can be combined 
	in the matrix  $C=g^{-1}\,L = g_r^{-1} \tt_\pm\,g_l^{-1}$ and, using the canonical coordinates \eqref{CC m=0}, one finds
	(see equation \eqref{C=} in Appendix \ref{app:SL2R})
	\be\label{C in pq}
		C=\begin{pmatrix} q_l p_r & -p_l p_r \\ -q_l q_r & p_l q_r\end{pmatrix}~.
	\ee
	Note that the conservation of $C=g^{-1}\,L$ follows from the equations of motion \eqref{Hamilton eq} and 
	from the nilpotency condition $L^2=0$, valid for the massless case.

	As a result, we obtain ten dynamical integrals given by quadratic combinations of four canonical variables $\{p_l, q_l, p_r, q_r\}$. The Poisson brackets of these functions  obviously form $\alg{sp}(4)$,
	\be\label{sp(4) generators}
		\alg{sp}(4) = \Span\left\{p_l^2\,,\, q_l^2\,,\, p_l\,q_l\,,~~ p_l p_r\,,\,p_l q_r\,,\,q_l p_r\,,\,q_l q_r\,,~~ p_r^2\,,\, q_r^2\,,\,p_r\,q_r\right\}~.
	\ee
	This algebra is isomorphic to $\so(2,3)$, which corresponds to the conformal symmetry of $\AdS_3$. 

	Usually, the standard form of the $\so(2,3)$ algebra is depicted as
	\be\label{so(2,3) matrix algebra}
	\{J_{AB}, J_{CD}\}=\eta_{AC}\,J_{BD}+\eta_{BD}\,J_{AC}
	-\eta_{AD}\,J_{BC}-\eta_{BC}\,J_{AD}~,
	\ee
	with $A,B.\ldots=0', 0, 1,2,3$ and $\eta_{AB} = \diag(-1,-1,1,1,1)$ being the metric tensor of $\Reals^{2,3}$. On the basis of 
	\eqref{AdS isometry tr}-\eqref{SL(2) conformal tr} one obtains
	\be\label{basis 1}
	J_{\m 0'}=\f 1 2(L_\m+R_\m)~, \quad J_{\m\n}=\f 1 2\,\eps_{\m\n}\,^\rho (L_\rho-R_\rho)~, \quad J_{30'}=-C_{0'}~,\quad
	J_{3\m}=C_\m~,
	\ee
	where $L_\m$, $R_\m$ are given by \eqref{DI m=0} and $C_{0'}, C_\m$, with $\m=0,1,2$, are obtained from \eqref{C in pq}.
	The canonical Poisson brackets $\{p_l,q_l\}=\{p_r,q_r\}=1$ indeed provide the algebra 
	\eqref{so(2,3) matrix algebra}.

	Quantum realization of these commutation relations is obtained by the Weyl ordering.
	Using the creation-annihilation operators, $a_l^\pm=\f{1}{\sqrt{2}}(p_l\pm iq_l)$ and
	$\,a_r^\pm=\f{1}{\sqrt{2}}(p_l\pm iq_l)$,  one obtains 
	the energy operator 
	\be\label{Energy Op}
	E=\f 1 2(a_l^+ a_l^- +a_r^+ a_r^-)+\f 1 2~,
	\ee
	with eigenstates $|n_l,n_r\ra$.
	The operators of the right sector $H_r=\f 1 2(a_r^+a_r^-+\f 1 2)$, 
	$J_r^\pm=\f 1 2 a_r^\pm\,^2$ realize the $\sls(2,\Reals)$ 
	algebra which contains two unitary irreducible representations, with minimal eigenvalues
	of $H_r$ equal to $1/4$ and to $3/4$. The first is realized on the even level eigenstates ($n_r=2k$) 
	and the second on the odd ones ($n_r=2k+1$).  
	The operators of the left sector $H_l=\f 1 2(a_l^+a_l^-+\f 1 2)$, 
	$J_l^\pm=\f 1 2 a_l^\pm\,^2$ give a similar representation of $\sls(2,\Reals)$. In addition,
	one has four $\alg{sp}(4)$ generators $a^-_l a^-_r$, $a^-_l a^+_r$, $a^+_l a^-_r$, $a^+_l a^+_r$.
	Since the symmetry generators are quadratic in creation-annihilation operators, they preserve the parity of $n_l+n_r$.
	Thus, the constructed representation of $\alg{sp}(4)$ splits in two irreducible representations, with even and odd $n_l+n_r$, 
	respectively.

	Note that the representation of $\sls(2,\Reals)$ given by \eqref{Hamiltonian}-\eqref{J_pm} at $m=0$ describes 
	the unitary irreducible representations either  with $\m=1/4$ or with $\m=3/4$ (see \eqref{E min}). 
	They correspond to the Neumann and Dirichlet boundary conditions of the oscillator eigenfunctions at $q=0$, respectively. 
	Therefore, in the limit $m\rightarrow 0$ one does not get the $\alg{sp}(4)$ symmetry and the case
	$m=0$ has to be treated separately.

\section{The \texorpdfstring{$\AdS_3$}{AdS3} superparticle}\label{sec:Super}

\subsection{Coset construction}\label{subsec:Coset}
	In the context of the AdS/CFT correspondence, a particularly interesting class of $\AdS$ string theories are the ones exhibiting classical integrability. Typically, these are formulated as sigma models on semi-symmetric spaces \cite{Zarembo:2010sg}, that is supercosets $G/H$ with $G$ containing the $\AdS_{N+1}$ isometry group $\SO(2,N)$ and its stabilizer $H$ containing $\SO(1,N)$.

	The case of $\AdS_3$ is somewhat special as here the cosets of interest take the form $G^2/H$ with $H$ the bosonic part of the diagonal subgroup of $G^2$, which is isomorphic to the bosonic subgroup of $G$. 
	Especially, in case of the $\AdS_3$/CFT$_2$, see {\it e.g.} \cite{Babichenko:2009dk}, the relevant coset is $\mathfrak{D}(2,1;\a)^2/\SO(1,2)\times\SO(3)^2$, with the special cases $\PSU(1,1|2)^2 / \SO(1,2) \times \SO(3)$ for $\alpha=0$ or $\alpha=1$ as well as $\OSP(4|2)^2/\SO(1,2)\times\SO(3)^2$ for $\alpha=1/2$. In this work we will instead study the simpler coset $\OSP(1|2)^2/\SLS(2;\Reals)$, which also has this feature.

	%Superparticle dynamics in AdS backgrounds can be described by a $G/H$  coset scheme, where $G$ is a supergroup which contains $\SO(2,N)$ as a bosonic subgroup and $H$ is associated with $\SO(1,N)$. This scheme is based on the representation of AdS$_{N+1}$ as the coset space $\SO(2,N)/\mbox{SO}(1,N)$. Since  $\so(2,2)=\mbox{sl}(2,\Reals)\oplus\mbox{sl}(2,\Reals)$, one can choose for $\AdS_3$ the supergroup $G=\OSP(1|2)\times\OSP(1|2)$ and the role of the gauge group $H$ plays $\SLS(2,\Reals)$, which acts symmetrically on both $\OSP(1|2)$ parts.

	But first, let us discuss the general case of a coset of the form $G^2/H$, where $H$ does not necessarily have to correspond to the bosonic subgroup of $G$. The group element $g\in G^2$ is given as the pair $g=(u,v)$ with $u\in G$ and $v\in G$ and the action of stabilizer subgroup $H\subset G$ on $G^2$ is defined by $(u,v)\mapsto (hu, hv)$, where $h\in H$.
	The Lie algebras of $G$ and $H$ are denoted by $\alg{g}$ and $\alg{h}$, respectively, and we introduce the orthogonal completion of $\alg{h}$ in $\alg{g}$, which is denoted by $\alg{h}_\perp$.
	The metric tensor on $\alg{h}$ is defined by a normalized Killing form $\rho_{ab}=\la \,\tt_a\,\tt_b\,\ra$ of basis vectors $\tt_a\in \alg{h}$, whereas the basis of $\alg{h}_\perp$ is denoted by ${\bf s}_\a$. It is easy to check that the quadratic form $\rho^{ab}\la\,\tt_a\,v\,\ra \la\,\tt_b\,v\,\ra$ with $v\in \alg{g}$ is invariant under the transformations $v\mapsto h v h^{-1}$ for any $h\in  H$.

	The superparticle action is then given in the coset scheme by
	\be\label{coset action}
	S=\int \dd \tau\,\left[\frac{\la \,\tt_a(\dot{u}\,u^{-1}-\dot{v}\,v^{-1})\,\ra
	\la \,{\bf t}^a(\dot{u}\,u^{-1}-\dot{v}\,v^{-1})\,\ra}{2e}
	-\frac{e m^2}{2}\right]~,
	\ee
	and it is invariant under  the gauge transformations 
	$u(\tau)\mapsto h(\tau) u(\tau)$,  $v(\tau)\mapsto h(\tau) v(\tau)$, 
	with $h(\tau)\in H$. 
	The Faddeev-Jackiw method provides the following first order action
	\be\ba\label{coset action 1}
	S=\int \dd \tau \Big[\la L_u\,\dot{u}\,u^{-1}\ra+ \la L_v\,\dot{v}\,v^{-1}\ra
	- \frac{e}{2}\left(\la L_u\,L_u\ra+m^2 \right)+\\
	\l^a\la\,\tt_a(L_u+L_v)\,\ra +\xi^\a_u\la\,{\bf s}_\a\,L_u\,\ra+
	\xi^\a_v\la\,{\bf s}_\a\,L_v\,\ra\Big]~,
	\ea\ee
	where $e$, $\l^a$, $\xi^\a_u$ and $\xi^\a_v$ are Lagrange multipliers and one obtains
	the constraints 
	\be\label{L=0}
	\la L_u\,L_u\ra+m^2=0~, \qquad L_u\in \alg{h}~, \qquad L_v=-L_u~.
	\ee
	The system is then described by the 1-form and the Noether charges
	\be\label{1-form and DI}
	\Th=\la L_u(\dd u\,u^{-1}-\dd v\,v^{-1})\,\ra~,\qquad R_u=u^{-1}L_u\,u~, \qquad R_v=-v^{-1}L_u\,v~.
	\ee
	Introducing gauge invariant variables $g=v^{-1}u$ and $L=v^{-1}L_u\,v$, from 
	\eqref{1-form and DI} we find
	\be\label{1-form and DI 1}
	\Th=\la\, L\,\dd g\,g^{-1}\,\ra~, \qquad  R_u=g^{-1}L\,g~, \qquad      R_v=-L~.
	\ee
	
	It is interesting to note that, in comparison, the superparticle action on the (super)group manifold $G$  \eqref{action 2} would yield $G$ orbits of some element $L_u$ of $\alg{g}$ instead of an element of its subalgebra $\alg{h}$. Hence, the action \eqref{coset action} on the coset $G^2/H$ corresponds to a subclass of orbits of the action \eqref{action 2} on the group manifold $G$.\footnote{The actions coincide, if in fact the mass-shell condition \eqref{mass-shell} requires $L_u$ to be an element of $\alg{h}$. This appears to happen for $\alg{h}$ being the bosonic subalgebra of $\alg{g}$, as is the case for $\OSP(1|2)^2/\SLS(2,\Reals)$, and the mass parameter taken to be pure body, $m\in\Reals$.}

	As mentioned above, we will be interested in the $\AdS_3$ superparticle corresponding to $G=\OSP(1|2)$ and $H=\SLS(2,\Reals)$. Then $L_u\in\mbox{sl}(2,\Reals)$ and $L$ is on its $\OSP(1|2)$ orbit. The bosonic case is given by $G=H=\SLS(2,\Reals)$, for which the reduction scheme describes a particle on $\SLS(2,\Reals)$ considered just in the previous section.

\subsection{Massive particle on \texorpdfstring{$\OSP(1|2)$}{OSP12}}

	First we introduce necessary notations and normalization in the $\osp(1|2)$ algebra.

	The standard basis of  $\osp(1|2)$ is given by the matrices
	\begin{align}\label{osp(1.2) basis}
		&\TT =\begin{pmatrix} 1 & 0 & 0\\ 0 & -1 & 0\\ 0 & 0 & 0 \end{pmatrix},\qquad\quad
			\TT_+=\begin{pmatrix} 0 & 1 & 0\\ 0 & 0 & 0\\ 0 & 0 & 0 \end{pmatrix},\qquad\quad
			\TT_-=\begin{pmatrix} 0 & 0& 0\\ 1 & 0 & 0\\ 0 & 0 & 0 \end{pmatrix},&\\
		\label{fermion basis}
		&\qquad\qquad\qquad{\bf S}_+=\begin{pmatrix} 0 & 0 & 1\\ 0 & 0 & 0\\ 0 & 1 & 0 \end{pmatrix},\qquad\qquad
			{\bf S}_-=\begin{pmatrix} 0 & 0 & 0\\ 0 & 0 & -1\\ 1 & 0 & 0\end{pmatrix},&
	\end{align}
	and they satisfy the commutation relations
	\be\ba\label{osp(1.2) algebra}
		&[\TT ,\,\TT_\pm]=\pm 2\TT_\pm~,& \qquad &[\TT_+,\,\TT_-]=\TT ~,& \\
		&[\TT ,\,{\bf S}_\pm]=\pm {\bf S}_\pm~,& &[\TT_\pm,\,{\bf S}_\mp]=- {\bf S}_\pm~,& \qquad
	[\TT_\pm,\,{\bf S}_\pm]=0~,\\
		&[{\bf S}_+ ,\,{\bf S}_-]_+=\TT ~,&  &[{\bf S}_\pm,\,{\bf S}_\pm]_+=\pm 2\TT_\pm~.& 
	\ea\ee

	The normalized supertrace  $\langle {\bf a}\, \bf b \rangle=
	\frac{1}{2}\big(({\bf a}\, {\bf b})_{11}+({\bf a}\, {\bf b})_{22}-
	({\bf a}\, {\bf b})_{33}\big)$ provides an inner product on $\osp(1|2)$
	with the following nonzero components
	\be\label{supertaces}
		\langle \TT \,\TT  \rangle=1~, \qquad \langle \TT_+\,\TT_- \rangle=\frac{1}{2}~,
		\qquad \langle {\bf S}_+\,{\bf S}_- \rangle=- \langle {\bf S}_-\,{\bf S}_+ \rangle=1~.
	\ee

	With this, we start from the action \eqref{coset action}, where $u$ and $v$ are group elements in $\OSP(1|2)$, the basis elements $\tt_a$ correspond to the bosonic generators \eqref{osp(1.2) basis} and $\la\, \cdot \,\ra$ denotes the normalized supertrace. 
	In the first order formalism one again gets the action \eqref{coset action 1} where now $L$ lies on the $\OSP(1|2)$ orbit of an element of the bosonic subalgebra $\sls(2,\Reals)$. 
	As for the purely bosonic particle in the last section, all that is left is to analyze the presymplectic form $\Th=\la L\,\dd g\,g^{-1}\ra$ and the Noether charges $L$ and $R=g^{-1}\,L\,g$ on the constrained surface $\la L\,L\ra+m^2=0$.

	In the massive case $L$ and $R$ are on the adjoint 
	orbit of $m\,\TT_0$, where $\TT_0=\TT_+-\TT_-$ is a unit timelike element of  $\osp (1|2)$. 
	Taking a parametrization similar to \eqref{L=t0,R=t0},
	\be\label{L=T0,R=T0}
		L=m\, g_l\,\TT_0\,g_l^{-1}~, \qquad R=m\, g_r^{-1}\,\TT_0\,g_r~, \qquad g=g_l\,g_r~,
	\ee
	splits the presymplectic form again into the left and right parts
	\be\label{SUSY 1-form m}
		\Th= \Th_l + \Th_r~,\qquad
		\Th_l = m\la\,\TT_0\,g_l^{-1}\,\dd g_l\,\ra~,\qquad
		\Th_r = m\la\,\TT_0\,\dd g_r\,g_r^{-1}\,\ra~.
	\ee

	For $g_l$ and $g_r$ let us take the parametrization
	\be\label{SUSY g_l-g_r m}
		g_l=e^{\g_l\,\TT_+}\,e^{\a_l\,\TT }\,e^{\zeta_l \,{\bf S}_+}\,e^{\eta_l\,{\bf S}_-}\,e^{\th_l\,\TT_0}~,\qquad
		g_r=e^{\th_r\,\TT_0}\,e^{\eta_r\,{\bf S}_-}\,e^{\zeta_r \,{\bf S}_+}\,
	\,e^{\a_r\,\TT }\,e^{\g_r\,\TT_+}~,
	\ee
	where $\eta_{l,r}$ and $\zeta_{l,r}$ are fermionic, {\it i.e.} Grassmann odd, parameters while the bosonic parameters $\th_{l,r}$, $\a_{l,r}$ and $\g_{l,r}$ correspond to Iwasawa type decomposition \eqref{Iwasawa}.
	Technical details of the parametrization \eqref{SUSY g_l-g_r m} are deferred to Appendix \ref{app:OSP}, where we also present some useful formulas. 

	Calculations of the Noether charges $L=m g_l\,\TT_0\,g_l^{-1}$ and   $R=mg_r^{-1}\,\TT_0\,g_r$ \eqref{L=T0,R=T0} as well as of the presymplectic forms $\Th_l = m\la\,\TT_0\,g_l^{-1}\,\dd g_l\,\ra$ and $\Th_r =  m\la\,\TT_0\,\dd g_r\,g_r^{-1}\,\ra$ \eqref{SUSY 1-form m} then yields
	\be\ba\label{SUSY L in angles}
		&L=m \begin{pmatrix}
			-\g_l\,e^{-2\a_l} & \g_l^2\,e^{-2\a_l}+e^{2\a_l}-2\,e^{2\a_l}\eta_l\zeta_l & 
			\g_l e^{-\a_l}\zeta_l + e^{\a_l}\eta_l\\ 
			-e^{-2\a_l} & \g_l\,e^{-2\a_l}  & e^{-\a_l}\zeta_l\,\\ 
			-e^{-\a_l}\zeta_l \,&  \g_l e^{-\a_l}\zeta_l + e^{\a_l}\eta_l  & 0
	\end{pmatrix},\\[1mm]
		&R=m \begin{pmatrix}
			\,\g_r\,e^{2\a_r} & \g_r^2\,e^{2\a_r}+e^{-2\a_r}-2\,e^{-2\a_r}\eta_r\zeta_r & 
			\g_r e^{\a_r}\zeta_r - e^{-\a_r}\eta_r\\ 
			\,-e^{2\a_r} & -\g_r\,e^{2\a_r}  & -e^{\a_r}\zeta_r\,\\ 
			\,e^{\a_r}\zeta_r \,&  \g_r e^{\a_r}\zeta_r - e^{-\a_r}\eta_r  & 0
	\end{pmatrix},
	\ea\ee
	\be\ba\label{1-form l}
		&\Th_l=\f{ m}{2}(\eta_l\dd \eta_l+\zeta_l\dd \zeta_l- e^{-2\a_l}\dd \g_l-2\dd \th_l)~,\\[1mm]
		&\Th_r=-\f{ m}{2}(\eta_r\dd \eta_r+\zeta_r\dd \zeta_r+e^{2\a_r}\dd \g_r+2\dd \th_r).
	\ea\ee
	Similarly to the bosonic case we introduce the variables 
	\be\ba\label{SUSY CC m}
		&p_l=\sqrt m\,\g_l\, e^{-\a_l},& ~~ &q_l=\sqrt m\, e^{-\a_l},& \quad
		&p_r=\sqrt m\,\g_r\, e^{\a_r},& ~~ &q_r=\sqrt m\, e^{\a_r},&\\[1mm]
		&\psi_l=\sqrt m\,\zeta_l\,e^{-i\,\f{\pi}{4}},&   &\chi_l=\sqrt m\,\eta_l\,e^{-i\,\f{\pi}{4}},& \quad
		&\psi_r=\sqrt m\,\zeta_r\,e^{i\,\f{\pi}{4}},&
		&\chi_r=\sqrt m\,\eta_r\,e^{i\,\f{\pi}{4}},&
	\ea\ee
	and obtain the canonical symplectic form
	\be\label{SUSY CF}
		\Omega = \dd \Th = \Omega_l + \Omega_r~,\qquad \Omega_l = \dd \Th_l = \f i 2(\dd \psi_l\wedge\dd \psi_l+\dd \chi_l\wedge\dd \chi_l)+\dd p_l\wedge\dd q_l~,
	\ee
	and similarly for $\Omega_r = \dd \Th_r$. %= \Omega_l\big|_{l\mapsto r}\,$.
	Suppressing indices, let us gather phase space variables of the left, respectively, right sector into $(2|2)$ vectors $\rho^a =(p,q,\psi,\chi)$, hence $\Omega = \frac{1}{2} \dd \rho^a \omega_{a b} \dd \rho^b$ with
	\be
		\omega_{a b} = \begin{pmatrix}
		                0&1&0&0 \\ -1&0&0&0\\ 0&0&i&0\\ 0&0&0&i
		               \end{pmatrix},\qquad\quad
		\omega^{a b} = (\omega_{a b})^{-1} = \begin{pmatrix}
		                0&-1&0&0 \\ 1&0&0&0\\ 0&0&-i&0\\ 0&0&0&-i
		               \end{pmatrix}=-\omega_{a b}~.
	\ee
	Up to an overall sign, this then determines the Poisson bracket of two functions $\calA$ and $\calB$ on phase space to take the form
	\be
		\{\calA,\calB\} = -\calA\,\frac{\overleftarrow{\p}}{\p \rho^a} \omega^{a b } \frac{\overrightarrow{\p}}{\p \rho^b} \calB = 
		%\frac{\p \calA}{\p p}\frac{\p \calB}{\p q} - \frac{\p \calA}{\p q}\frac{\p \calB}{\p p} 
		\calA \left(\overleftarrow{\p_p} \overrightarrow{\p_q} - \overleftarrow{\p_q} \overrightarrow{\p_p}%\right) \calB
		%+ i \calA \left(
		+i\,\overleftarrow{\p_{\psi}} \overrightarrow{\p_{\psi}} +i\, \overleftarrow{\p_\chi} \overrightarrow{\p_\chi}\right) \calB~.
	\ee
	In particular, this yields the non-vanishing Poisson brackets
	\be\label{C-PB}
		\{p_l, q_l\}=1~,\quad \{\psi_l, \psi_l\}=\{\chi_l, \chi_l\}=i~,\qquad
			\{p_r, q_r\}=1~, \quad \{\psi_r, \psi_r\}=\{\chi_r, \chi_r\}=i~,
	\ee
	and for example $\{i\chi\,\psi, \psi\}=-\chi$ and  $\{i\chi\,\psi, \chi\}=\psi$. The odd variables $\psi_{l,r}$ and $\chi_{l,r}$ are real and we construct the standard fermionic creation-annihilation variables by\,\footnote{We use the down $\pm$ indices for 
	'chiral' components and the  upper $\pm$ indices for complex coordinates.}  
	\be\label{f-f*}
		f_l^\pm=\f{\psi_l\pm i\chi_l}{\sqrt 2}~, \qquad\qquad
		f_r^\pm=\f{\psi_r\pm i\chi_r}{\sqrt 2}~,
	\ee
	hence $\{f^\pm_l,f^\mp_l\}=\{f^\pm_r,f^\mp_r\}=i$ and all other vanishing. Note that $i\chi\psi=f^+\,f^-$ is also real.

	In terms of the canonical variables \eqref{SUSY CC m} the Noether charges become
	\be\ba\label{SUSY L in CC}
		&L=\begin{pmatrix}
			-p_l\,q_l & p_l^2+m^2 q_l^{-2}-2 i m\,q_l^{-2}\,\chi_l\psi_l & 
			(p_l  \psi_l + m q_l^{-1}\,\chi_l)e^{i\,\f{\pi}{4}}\\ 
			-q^2_l & p_l\,q_l  & q_l\psi_l\,e^{i\,\f{\pi}{4}}\\ 
			-q_l\psi_l\,e^{i\,\f{\pi}{4}}\,& (p_l  \psi_l + m q_l^{-1}\,\chi_l)e^{i\,\f{\pi}{4}} & 0
		\end{pmatrix},\\[2mm]
		&R=\begin{pmatrix}
			p_r\,q_r & p_r^2+m^2 q_r^{-2}+2im\,q_r^{-2}\,\chi_r\psi_r & (p_r  \psi_r + m q_r^{-1}\,\chi_r)e^{-i\,\f{\pi}{4}}\\ 
			-q^2_r & -p_r\,q_r  & -q_r\psi_r\,e^{-i\,\f{\pi}{4}}\\ 
			q_r\psi_r\,e^{-i\,\f{\pi}{4}}\,& (p_l  \psi_r + m q_r^{-1}\,\chi_r)e^{-i\,\f{\pi}{4}}  & 0
		\end{pmatrix}.
	\ea\ee

	Introducing the dynamical integrals related to the Noether charges
	\be\ba\label{SUSY DI-r}
		&L_2=\la \TT \,L \ra~, ~~~ L_\pm=\la \TT_\pm\,L \ra~, \qquad
			&&R_2=\la \TT \,R \ra, ~~~ R_\pm=\la \TT_\pm\,R \ra~ ,\\[2mm]
		&l_\pm=\la {\bf S}_\pm\,L \ra\,e^{-i\f{\pi}{4}}~,
			&&r_\pm=\la {\bf S}_\pm\,R \ra\,e^{i\f{\pi}{4}}
	\ea\ee
	from \eqref{SUSY L in CC} we find
	\be\ba\label{osp DI}
		&L_2=-p_l\,q_l~, ~~\quad L_+=-\f 1 2\, q_l^2~, \qquad &&R_2=p_r\,q_r~, \quad R_+=-\f 1 2\, q_r^2~, \\
		&L_-=\f 1 2\left(p_l^2+\f{m^2}{q_l^2}\right)+i\,\f{m}{q_l^2}\,\,\chi_l\,\psi_l~,  
		&&R_-=\f 1 2\left(p_r^2+\f{m^2}{q_r^2}\right)+i\,\f{m}{q_r^2}\,\,\chi_r\,\psi_r~,\\
		&l_+= q_l\psi_l~, \quad     l_-=\f{m}{q_l}\,\chi_l-p_l\,\psi_l~,
		&&r_+= q_r\psi_r~,  ~~~\quad       r_-=\f{m}{q_r}\,\chi_r-p_r\,\psi_r~.
	\ea\ee
	The Poisson brackets of the right functions form the algebra
	\be\ba\label{PB osp DI}
		&\{R_2, R_\pm\}=\pm 2R_\pm~, \qquad&&\{R_+, R_-\}=R_{\,2}~,&&\\
		&\{R_2, r_\pm\}=\pm r_\pm~, &&\{R_\pm, r_\mp\}=-r_\pm~,\qquad&&\{R_\pm, r_\pm\}=0~,\\
		&\{r_+, r_-\}=-iR_{2}~, &&\{r_\pm, r_\pm\}=\mp 2iR_\pm~,&& 
	\ea\ee
	which is equivalent to the commutation relations of the basis elements \eqref{osp(1.2) algebra} with the replacements ${\bf S}_\pm \mapsto {\bf S}_\pm \,e^{-i\f \pi 4}$.
	The Poisson brackets of the left functions form the same algebra up to a sign, as in \eqref{PB L-R}. Therefore, due to similarity of the left and right sectors, in the following let us focus on the right sector and drop the corresponding index $r$.

	To pass to the quantum theory we apply the usual canonical quantization rule
	\be\label{SUSY CC}
		[q, p]=i~, \qquad [\chi, \chi]_+=[\psi, \psi]_+=1~, \qquad [\chi, \psi]_+=0~.
	\ee

	The quantum version of the symmetry generators are then obtained from the classical expressions \eqref{osp DI}. As in the purely bosonic case, see above \eqref{Hamiltonian}, by this only $R_2$ (and $L_2$) exhibit ambiguous operator ordering.
	Choosing again the Weyl ordering and the coordinate representation, we get $R_2=-i(q\p_{p}+1/2)$.

	Computation of the commutation relations then yields
	\be\ba\label{CR of osp DI}
		&[R_2, R_\pm]=\mp 2i\,R_\pm~,& \qquad &[R_+, R_-]=-i\,R_{\,2}~,&\\
		&[R_2, r_\pm]=\mp i\,r_\pm~,& &[R_\pm, r_\mp]=i\,r_\pm~,&   ~~~~~~[R_\pm, r_\pm]=0~,\\
		&[r_+, r_-]_+=- R_{2}~,&          &[r_\pm, r_\pm]_+=\mp 2 R_\pm~,& 
	\ea\ee
	which is the quantum version of \eqref{PB osp DI} in compliance with the rule $\{\calA,\calB\}\mapsto i[\calA,\calB]_\pm$.

	In terms of the fermionic creation and annihilation operators $(f^+, f^-)$ introduced in \eqref{f-f*} one gets $i\chi \psi=f^+\,f^--1/2$ and the (right) energy operator $H =\f 1 2(R_- -R_+)$ becomes 
	\be\label{osp H}
	H=\f 1 4\left(-\p^2_q+q^2+\f{m^2}{q^2}+\f{m}{q^2}(2f^+f^- -1)\right)~.
	\ee

	The canonical anti-commutation relations in \eqref{SUSY CC} are equivalent to
	\be\label{ff}
	[f^-, f^-]_+=[f^+, f^+]_+=0~, \qquad [f^-,f^+]_+=1~,
	\ee
	which is realized in the space of two component spinors and one gets 
	\be\label{spinor WF}
		f^-=\begin{pmatrix} 0&1\\0&0 \end{pmatrix}, \qquad 
		f^+=\begin{pmatrix} 0&0\\1&0 \end{pmatrix}, \qquad
		H=\begin{pmatrix} H_0&0\\0&H_1 \end{pmatrix},
	\ee
	with
	\be\label{SUSY Hamiltonian}
		H_0=\f 1 4\left(-\p^2_q+q^2+\f{m^2-m}{q^2}\right)~, \qquad  
		H_1=\f 1 4\left(-\p^2_q+q^2+\f{m^2+m}{q^2}\right)~.
	\ee
	The Hamiltonians $H_0$ and $H_1$ have the oscillator spectrum with minimal eigenvalues $\m_0=\f{2m+1}{4}$ and $\m_1=\f{2m+3}{4}$, respectively, and they are represented in the form
	of supersymmetric quantum mechanics \cite{Cooper:1994eh}
	\be\label{SUSY Hamiltonians}
		H_0=A^+ A^-+\f{2m+1}{4}~, \qquad\qquad H_1=A^-A^+ +\f{2m-1}{4}~,
	\ee
	with 
	\be\label{Apm}
		A^+=\f 1 2 \left(q-\f m q-\p_q\right)~, \qquad\qquad A^-=\f 1 2 \left(q-\f m q+\p_q\right)~.
	\ee

	Introducing the rising-lowering operators for the Hamiltonian $H$ 
	\be\label{Jpm-jpm}
		J^\pm=\f 1 2(R_++R_-\pm iR_2)~, \qquad 
		j^\pm=\f {1}{\sqrt{2}}(r_+  \pm ir_-)~,\\[2mm]
	\ee
	one gets the following form of the $\osp(1|2)$ algebra \eqref{CR of osp DI}
	\be\ba\label{osp new}
		&[H, J^\pm]=\pm J^\pm,& \quad   &[H, j^\pm]=\pm\f 1 2\, j^\pm,& \quad 
		&[j^\pm, j^\pm]_+=-2J^\pm,& \quad [j^+,j^-]_+=2H,\\[2mm]
		&[J^-,J^+]=2H,&    &[J^\pm, j^\mp]=\pm j^\pm,&      &[J^\pm, j^\pm]=0~.&
	\ea\ee

	This representation of $\osp(1|2)$ is unitary equivalent to the Holstein-Primakoff type representation given by \cite{Heinze:2015oha}
	\be\label{b,f to B,F}
		E=\m+b^\dag b+\frac{f^\dag f}{2}~, \quad
		B=\sqrt{2\m+b^\dag b+f^\dag f}\,\,b~,   \quad  
		F=\sqrt{2\m+b^\dag b}\,f+f^\dag\,b~,
	\ee
	together with $B^\dag$ and $F^\dag$. Here, $\m=\frac{2 m + 1}{4}$, $b$ and $b^\dag$ as well as $f$ and $f^\dag$ are the standard bosonic and fermionic creation-annihilation 
	operators, $[b,b^\dag]=1$ and $[f,f^\dag]_+=1$, and
	the unitary map to \eqref{b,f to B,F} is provided by $\{E,B,B^\dag,F,F^\dag\} \leftrightarrow \{H,-J^-,-J^+,j^-,j^+\}$.

%$E \leftrightarrow H$, $\,B \leftrightarrow  -J^-$, 
%$\,B^\dag \leftrightarrow  -J^+$, 
%$\,F \leftrightarrow  j^-$, $\,F^\dag \leftrightarrow  j^+$.

	Let us establish the corresponding canonical transformation at the classical level.
	Using \eqref{f-f*} and \eqref{osp DI}, the dynamical integrals $H$, $J^\pm$, $j^\pm$ can be written as
	\bea\label{H-Jpm}
		&&H=\f 1 4\left(p^2+q^2 +{{\tilde m}^2}/{q^2}\right)~, \qquad 
		J^\pm=\f 1 4\left(p^2-q^2+{{\tilde m}^2}/{q^2}\right)\pm \f i 2\,\, pq\\
		\label{jpm}
		&&j^\pm=\f 1 2(q+m/q \mp ip)f^\pm +\f 1 2(q-m/q \mp ip)f^\mp~,
	\eea
	with $\tilde m=m+f^+f^-$ and one gets $j^+j^-=mf^+f^-$. 
		Similarly,  $F^*F=mf^*f$, as it follows
	from the classical form of \eqref{b,f to B,F}
	\be\label{b,f to B,F cl}
		E=\f{\hat m}{2}+b^* b~, \quad
		B=\sqrt{\hat m+b^* b}\,\,b~,   \quad  
		F=\sqrt{m+b^* b}\,f+f^*\,b~,
	\ee
	where $\hat m=m+f^*f$.  Both sets of generators then have the same Casimir 
	\be\label{Casimir}
		H^2-J^+J^--\f 1 2\,j^+j^-=\f{m^2}{4}~, \qquad\qquad E^2-B^*B-\f 1 2 F^*F=\f{m^2}{4}~.
	\ee

	We use the equations 
	\be\label{E=H}
		H=E, \quad J^+=-B^*, \quad J^-=-B~,\quad j^+=F^*,\quad j^-=F~,
	\ee
	to find the canonical map between the variables $(p,q,f^\pm)$ and $(b^*,b,f^*,f)$.

	Note that the odd part of \eqref{E=H} implies $f^+f^-=f^*f$, hence $\tilde m=\hat m$. Then, from \eqref{H-Jpm}
	\be\label{H -Jmp}
		q^2\pm ipq=2(H-J^\mp)~,
	\ee
	and by the bosonic part of \eqref{E=H} one finds
	\be\label{p-q=}
		q=\sqrt{2E+B^*+B}~, \qquad pq=i(B^*-B)~.
	\ee
	Using again \eqref{H -Jmp} and the odd part of \eqref{E=H}, we obtain
	\be\label{fpm=}
		f^+=\f{E+B+m/2}{\sqrt{(E+m/2)(2E+B^*+B)}}\,f^*, 
	\ee
	and $f^-$ is its complex conjugated. 

	Since $f^*f^*=0$,
	one can replace $\hat m$ by $m$ in the expressions of $E$, $B^*$, $B$ standing 
	in the right hand side of \eqref{fpm=}.\,\footnote{ By the same reason
	we use $m$ instead of $\hat m$ in the odd element of \eqref{b,f to B,F cl}.}
	After this replacement, the bosonic  factor in \eqref{fpm=}
	gets unit norm, which helps to check that the transformation \eqref{p-q=}-\eqref{fpm=}
	from $(b^*,b,f^*,f)$ to $(p,q,f^\pm)$  is indeed canonical   
	\be\label{C-Tr}
		\dd p\wedge\dd q +i\dd f^+\wedge\dd f^-=i\dd b^*\wedge\dd b+i\dd f^*\wedge\dd f~.
	\ee

	One can repeat the same for the right part of the system and obtain a parameterization of all dynamical integrals
	in terms of bosonic and fermionic oscillator variables.

\subsection{Massless particle on \texorpdfstring{$\OSP(1|2)$}{OSP12}}

	In the massless case $L$ and $R$ are on the adjoint orbit of the nilpotent element $\TT_+$
	\be\label{SUSY orbits m=0}
		L=g_l\,\TT_+\,g_l^{-1}~, \qquad R=g_r^{-1}\,\TT_+\,g_r~.
	\ee
	Here we use the parametrization (see Appendix \ref{app:OSP})
	\be\label{SUSY gl,gr}
		g_l=e^{\th_l\,\TT_0}\,e^{\a_l\,\TT }\,e^{\zeta_l\,{\bf S}_-}\,
		e^{\eta_l\,{\bf S}_+}\,e^{\g_l\,\TT_+}~, \qquad\quad
		g_r=e^{\g_r\,\TT_+}\,e^{\eta_r\,{\bf S}_+}\,e^{\zeta_r\,{\bf S}_-}\,
		e^{\a_r\,\TT }\,e^{\th_r\,\TT_0}~,
	\ee
	which yields the Noether charges 
	\begin{align}\label{SUSY m=0 L in angles}
		&L=\begin{pmatrix}
		e^{2\a_l}\cos\th_l\,\sin\th_l & e^{2\a_l}\cos^2\th_l & e^{\a_l}\cos\th_l \,\,\zeta_l \\ 
		-e^{2\a_l}\sin^2\th_l & -e^{2\a_l}\cos\th_l\,\sin\th_l  & -e^{\a_l}\sin\th_l\,\,\zeta_l\,\\ 
		e^{\a_l}\sin\th_l\,\,\zeta_l & e^{\a_l}\cos\th_l\,\, \zeta_l & 0
		\end{pmatrix},\\
		\label{SUSY m=0 R in angles}
		&R=\begin{pmatrix}
			-e^{-2\a_r}\cos\th_r\,\sin\th_r & e^{-2\a_r}\cos^2\th_r & -e^{-\a_r}\cos\th_r \,\,\zeta_r \\ 
			-e^{-2\a_r}\sin^2\th_r & e^{-2\a_r}\cos\th_r\,\sin\th_r  & -e^{-\a_r}\sin\th_r\,\,\zeta_r\,\\ 
			e^{-\a_r}\sin\th_r\,\,\zeta_r & -e^{-\a_r}\cos\th_r \,\,\zeta_r & 0
		\end{pmatrix}.
	\end{align}
	The  presymplectic form is again given as the sum of the left and right parts 
	$\Th =\Th_l+\Th_r$, with
	$\Th_l=\la \TT_+\,g_l^{-1}\,\dd g_l\ra$ and 
	$\Th_r=\la \TT_+\,\dd g_r\,g_r^{-1}\ra$. Using again \eqref{SUSY gl,gr}, one finds
	\be\label{SUSY 1-forms m=0}
	\Th_l=
	\f 1 2(\zeta_l\,\dd \zeta_l-e^{2\a_l}\dd \th_l)~, \qquad \Th_r=
	-\f 1 2(\zeta_r\,\dd \zeta_r+e^{-2\a_r}\dd \th_r)~.
	\ee
	The Noether charges and the symplectic form do not depend on the odd variables 
	$(\eta_l,\,\eta_r)$, which reflects the $\k$-symmetry of the massless case.

	Similarly to \eqref{CC m=0}, canonical variables here are introduced by
	\be\label{SUSY CC m=0}
		p_l- i\,q_l = e^{\a_l} e^{i\,\th_l}~,\quad 
			\psi_l=\zeta_l \, e^{-i\,\f \pi 4}~, \qquad\quad
			p_r- i\,q_r = -e^{-\a_r} e^{i\,\th_r}~,\quad
			\psi_r=\zeta_r \, e^{i\,\f \pi 4}~,
		%&p_l=e^{\a_l}\,\cos\th_l,& ~~~ q_l=-e^{\a_l}\,\sin\th_l,   \qquad &p_r=-e^{-\a_r}\,\cos\th_r,& ~~~q_r=e^{-\a_r}\,\sin\th_r,\\[2mm]
		%& \psi_l=\zeta_l \, e^{-i\,\f \pi 4},&   &\psi_r=\zeta_r \, e^{i\,\f \pi 4},&
	\ee
	and one obtains
	\be\label{SUSY m=0 L in CC}
		L=\begin{pmatrix}
		-p_l\,q_l & p_l^2 & p_l \,\zeta_l \\ 
		-q_l^2 & p_l\,q_l & q_l\,\zeta_l\,\\ 
		-q_l\,\zeta_l & p_l\,\zeta_l & 0
		\end{pmatrix},
		\qquad\qquad
		R=\begin{pmatrix}
		p_r\,q_r & p^2_r & p_r \,\zeta_r \\ 
		-q^2_r & -p_r\,q_r  & -q_r\,\zeta_r\,\\ 
		q_r\,\zeta_r & p_r\,\zeta_r & 0
		\end{pmatrix},
	\ee
	\be\label{SUSY CF m=0}
	\dd \Th=\f i 2(\dd \psi_l\wedge\dd \psi_l+\dd \psi_r\wedge\dd \psi_r)+
	\dd p_l\wedge\dd q_l+\dd 
	p_r\wedge\dd q_r.
	\ee

	Hence, following \eqref{SUSY DI-r} we obtain ten dynamical integrals corresponding to the isometry group  $\OSP_l(1|2) \oplus\OSP_r(1|2)$, which take the simple form
	\be\label{osp DI m=0}
		p_l^2\,,~~q_l^2\,,~~p_l\,q_l\,,~~p_l\,\psi_l\,, ~~q_l\psi_l\,,~~\qquad
		p_r^2\,,~~ q_r^2\,,~~ p_r\,q_r\,, ~~  p_r\psi_r\,,~~q_r\,\psi_r\,.
	\ee
	Due to the masslessness, the symmetry algebra extends by nine additional dynamical integrals,
	\be\label{susy DI m=0}
		p_l p_r\,,~~~p_l q_r\,,~~~q_l p_r\,,~~~q_l q_r\,,\qquad 
		p_l \psi_r\,, ~~ q_l\psi_r\,, ~~ \psi_l p_r \,, ~~ \psi_lq_r\,, \qquad i\psi_l\psi_r~.
	\ee
	The leftmost four correspond to the $\SLS(2,\Reals)$ conformal transformations, while the remaining five follow from closure of the algebra.
	Altogether, the 19 functions in \eqref{osp DI m=0} and \eqref{susy DI m=0} comprise all possible real quadratic combinations of phase space variables and form the algebra $\osp(2|4)$. 
	This result matches with the supertwistor representation for the massless superparticle in three dimensional flat space, see for example \cite{Bengtsson:1987si}, because at least locally conformal theories do not distinguish between flat and $\AdS$ backgrounds.

	Quantization of the model is then straightforward. Similarly to \eqref{Energy Op}, for the bosonic variables we define creation-annihilation operators $a_l^\pm=\f{1}{\sqrt{2}}(p_l\pm iq_l)$ and $a_r^\pm=\f{1}{\sqrt{2}}(p_r\pm i q_r)\,$.
	Concerning the fermionic variables, it is crucial that after $\kappa$-symmetry we are still left with two real fermions, $\psi_l$ in the left and $\psi_r$ in the right sector, which is just enough to form one fermionic oscillator $\psi = \frac{1}{\sqrt{2}}(\psi_l +i\,\psi_r)$.\footnote{In comparison, recall that for the massless $\AdS_2$ superparticle on $\OSP(1|2)/\SO(2)$ \cite{Heinze:2015oha} $\kappa$-symmetry reduced the phase space to only one real fermion, which is insufficient for quantization of the model.} Following the canonical quantization rule $\{\calA,\calB\}\mapsto i[\calA,\calB]_\pm$ and adopting the ordering in \eqref{Energy Op} then yields a quantum realization of $\osp(2|4)$.

	Finally, we would like to note that at the classical level there is another attractive extension of the $\osp_l(1|2) \oplus\osp_r(1|2)$ algebra. Recall that the isometry algebra of the bosonic $\AdS_2$ particle on $\SLS(2,\Reals)/\SO(1,1)$ consists out of only one $\sls(2,\Reals)$, see also \cite{Heinze:2015oha}. % and \cite{Heinze:2016b}.
	For the massless case, this symmetry extends to the corresponding conformal symmetry of $\AdS_2$, which is the infinite dimensional Virasoro algebra $\alg{Vir}$, viz. the Witt algebra at the classical level.
	Moreover, it is well known that $\osp(1|2)$ is a subalgebra of the super Virasoro algebra in the NS sector $\alg{sVir}_{\rm NS}$. Therefore, it is tempting to extend the $\osp_l(1|2) \oplus\osp_r(1|2)$ algebra of the present massless $\AdS_3$ superparticle to a double copy of the classical super Virasoro algebra in the NS-NS sector, $\alg{sVir}_{{\rm NS},l}\oplus\alg{sVir}_{{\rm NS},r}$.	

	Focusing again on the right sector and suppressing once more the index $r$, for this we again introduce the Hamiltonian $H = \la R \TT_0 \ra$ \eqref{osp H} and raising-lowering functions \eqref{Jpm-jpm}. By \eqref{SUSY m=0 L in CC}, these become
	\be
		\label{H-Jpm m=0}
		H = \frac{1}{4}(p^2 +q^2)~,\qquad\quad 
		J^\pm = \frac{1}{4}(p \pm i\,q)^2~,\qquad\quad 
		j^\pm = \frac{-i}{\sqrt{2}}(p \pm i\,q) \psi~,
	\ee
	which resembles the $m\rightarrow 0$ limit of \eqref{H-Jpm} and which fulfill the classical version of \eqref{osp new}. Introducing the angle $\phi$ conjugate to $H$ as
	\be
		\phi = \text{arg}\left(J^+\right)~,\qquad e^{i\,\phi} = \frac{J_+}{H} =\frac{(p+iq)^2}{p^2+q^2}~,\qquad \{H,e^{i\,n\,\phi} \}= i\,n\,e^{i\,n\,\phi}~,
	\ee
	we get $J^\pm = H\,e^{\pm i\,\phi}$ and $j^\pm = \sqrt{2\,J^\pm}\,\psi = \sqrt{2\, H}\,e^{\pm \frac{i}{2}\,\phi}\,$. From this we guess the charges
	\be
		J_n =  H\,e^{\pm i\,n\,\phi}~,\qquad\qquad j_s = -i\sqrt{2\,J_{2\,s}}\,\psi = -i\sqrt{2\, H}\,e^{i\,s\,\phi}\,\psi~,
	\ee
	for $n\in\Integers$ and $s$ being half-integer, $s\in\Integers + \frac{1}{2}$, especially $J_0 = H$, $J_{\pm1} = J^\pm$ and $j_{\pm1/2} = j^\pm$.
	These charges indeed fulfill the classical super Virasoro algebra in the NS sector,
	\be
		\{J_m,J_n\}=-i(m-n)J_{m+n}~,\quad 
		\{J_m,j_s\}=-i\Big(\frac{m}{2}-s\Big)j_{m+s}~,\quad
		\{j_s,j_t\}=-2i\,J_{s+t}~.
	\ee

	We have not been able to quantize this classical representation of $\alg{sVir}_{{\rm NS},l}\oplus\alg{sVir}_{{\rm NS},r}$. This comes to no surprise as even quantization of the Virasoro algebra for the bosonic $\AdS_2$ particle is still an open question. %\cite{Heinze:2016b}.

	Furthermore, it has to be pointed out that in contrast to the extension of $\osp_l(1|2) \oplus\osp_r(1|2)$ superisometries to the superconformal algebra $\osp(2|4)$ the above extension to $\alg{sVir}_{{\rm NS},l}\oplus\alg{sVir}_{{\rm NS},r}$ does not actually correspond to a symmetry of the $\AdS_3$ superparticle action. However, it is known that the $\AdS_3$/CFT${}_2$ on $\AdS_3 \times \Sph^3 \times \mathrm{M}_4$ enjoys a {\it small}, respectively, {\it large} $\calN = (4,4)$ superconformal algebra, see {\it e.g.} \cite{Giveon:1998ns, Elitzur:1998mm, deBoer:1999gea, Gukov:2004ym} as well as the more recent works \cite{Gaberdiel:2013vva, Tong:2014yna}, which have $\alg{sVir}_{{\rm NS},l}\oplus\alg{sVir}_{{\rm NS},r}$ subalgebras. Hence, the discussed extension might show relevant for effective descriptions of string and even higher spin states.

\section{Conclusions}\label{sec:concl}
	Quantization of the Green-Schwarz superstring on $\AdS$ superspaces from first principles is still an open problem. To attain a better understanding, the work \cite{Frolov:2013lva} suggested to study orbit method quantization of semi-classical string solutions, where we have explored this idea in \cite{Heinze:2014cga} and \cite{Heinze:2015oha}. In this work, we continued this program and applied superisometry group orbit quantization to the $\k$-symmetric $\AdS_3$ superparticle on the coset $\OSP(1|2)^2/\SLS(2,\Reals)$.

	First, we reviewed how the method applies to bosonic $\AdS_3$ on the group manifold $\SLS(2,\Reals)$. The massive particle is described by orbits of a temporal $\sls(2,\Reals)$ element, with its norm given by the mass $m$, while for the massless case one has to consider orbits of a lightlike $\sls(2,\Reals)$ element. For both cases the calculations split up into left and right chiral sectors and the physical phase space of each sector is two dimensional, being a half-plane for the massive case while a full plane without origin for the massless case.
	From the left and right Noether currents we read off the dynamical integrals, where at the classical level the massless charges can formally be viewed as the $m \rightarrow 0$ limit of the massive charges. These fulfill the isometry algebra $\sls_l(2,\Reals)\oplus\sls_r(2,\Reals) \cong \so(2,2)$, which determined quantization of the system, yielding a quantum realization unitarily equivalent to the Holstein-Primakoff representation \cite{Dzhordzhadze:1994np}, see also \cite{Heinze:2014cga, Jorjadze:2015}. For the massless case we then observed how the isometries extend to the $\AdS_3$ conformal symmetries $\alg{sp}(4) \cong \so(2,3)$.

	Next, we turned to the $\AdS_3$ superparticle. For this we first discussed the superparticle action on cosets of the form $G^2/H$ and pointed out that generally in comparison to the superparticle action on $G$ it amounts to a subclass of orbits. In particular, focusing then on $G^2/H = \OSP(1|2)^2/\SLS(2,\Reals)$, the massive and massless particle are described by $\OSP(1|2)$ orbits of timelike and lightlike elements of $\sls(2,\Reals)$, respectively.
	Again, the calculation split into left and right chiral sectors, where apart from two real bosons the physical phase space of each sector contains two real fermions in the massive case whilst only one real fermion for the massless case. As anticipated \cite{Heinze:2015oha}, the latter reflects the underlying $\vk$-symmetry for the massless superparticle and importantly the two remaining real fermions could be combined into a fermionic oscillator, which can be quantized. The dynamical integrals respected the $\osp_l(1|2) \oplus \osp_r(1|2)$ super isometry algebra and quantization amounted to two copies of Holstein-Primakoff type quantum representations of $\osp(1|2)$ \cite{Heinze:2015oha}. For the massive case we observed that each chiral sector corresponds to the superparticle with superpotential $W=q-\f{2\m-1/2}{q}\,$ \cite{Cooper:1994eh}.
	For the massless case we demonstrated how the superisometries extend to the corresponding superconformal algebra $\osp(2|4)$. We finally pointed out that, at least at the classical level, there is another interesting extension of the superisometry algebra $\osp_l(1|2) \oplus \osp_r(1|2)$ to a double copy of the super Virasoro algebra in the NS sector, $\alg{sVir}_{{\rm NS},l}\oplus\alg{sVir}_{{\rm NS},r}\,$.

	Our work offers several future directions of research. As this article discusses orbit method quantization for what arguably amounts to the simplest quantum consistent massless $\AdS$ superparticle, a natural next step is to investigate $\AdS$ superparticles with a higher amount of supersymmetry, in particular the $\AdS_2$  and $\AdS_3$ superparticles build on the superalgebras $\su(1,1|1)$, $\psu(1,1|2)$, and $\alg{d}(2,1;\alpha)$, see also \cite{Bellucci:2002va, Ivanov:2002tb, Galajinsky:2010zy, Galajinsky:2011xp, Bellucci:2011hk, Orekhov:2014xra} and \cite{Kozyrev:2016mlo, Kozyrev:2013vla, Krivonos:2010zy, Galajinsky:2016wuc} and references therein. 
	
	In a longer term we would like to utilize this quantization scheme to the $\AdS_5 \times \Sph^5$ superparticle, see also \cite{Horigane:2009qb} and the recent work \cite{Arvanitakis:2016vnp}. Hence, apart from increasing the amount of supersymmetry another intermediate goal is to raise the dimension of the $\AdS$ space. Indeed, the found charges forming a quantum realization of $\sls_l(2,\Reals)\oplus \sls_r(2,\Reals)$, respectively, $\osp_l(1|2)\oplus \osp_r(1|2)$ can be rewritten in an $\so(2,2)$ scheme. 
	By this, the expressions become covariant under the $\so(2)\subset\so(2,2)$ corresponding to the rotations of the spatial directions of $\Reals^{2,2}$ embedding space. As we show in \cite{Heinze:2016lxs}, generalization of the $\so(2)$ to an $\so(N)$ covariance then yields an ansatz for quantum prescription of the bosonic $\AdS_{N+1}$ particle, respectively, the $\calN=1$ $AdS_{N+1}$ superparticle. A similar idea has been adopted in \cite{Galajinsky:2016wuc}, where the dynamical realization on $\SU(1,1|2)$ has been generalized to $\SU(1,1|N)$. 

	Another direction is to apply the orbit method to honest string solution, {\it viz.} ones storing more information than only the particle degrees of freedom. As advocated previously \cite{Frolov:2013lva}, we hope that such orbits open a window into computation of the string spectrum from first principles, especially for short strings. In particular, in view of \cite{Heinze:2014cga} we expect that the step from the superparticle on for example $\frac{\PSU(1,1|2)}{\SO(1,1)\times\SO(2)}\times {\rm T}^4$ to orbit quantization of more involved string solutions on this background should be manageable, thus yielding results for the spectral problem in the $\AdS_3$/CFT${}_2$ \cite{Babichenko:2009dk, David:2008yk, OhlssonSax:2011ms, Sundin:2012gc, Cagnazzo:2012se, Borsato:2012ud, Borsato:2013qpa}.

	Furthermore, we are also curious if our results may find application for the non-critical $\AdS_3$ superstring, see  also \cite{Hikida:2007sz}, and even of the $\AdS_3$ higher spin theory \cite{Vasiliev:1995dn, Vasiliev:1999ba}. Even for the non-critical string in bosonic $\AdS_3$ \cite{Brown:1986nw,Maldacena:2000hw} it seems promising to apply the orbit method as the model resembles the WZNW model on $\SLS(2,\Reals)$ \cite{Alekseev:1988ce}.

	Finally, lately there has been considerable interest in the so-called $\eta$-deformation \cite{Delduc:2013qra, Delduc:2014kha, Arutyunov:2013ega}, which amounts to a one-parameter integrable deformation of the of the $\AdS_5 \times \Sph^5$ superstring. However, even the particle dynamics on this background are still an open problem, see also \cite{Arutyunov:2016ysi}.
	For the truncation to $(\Sph^2)_\eta$, corresponding to the Fateev sausage model, geodesic motion has been solved recently \cite{Arutyunov:2016kve} but the non-closure of the orbits seems to stem a fundamental obstacle to quantization of this system. Also here the Kirillov-Kostant-Souriau method of coadjoint orbits might lead the way out, as its extension to quantum groups has been investigated \cite{KirillovMerits}.

\subsection*{Acknowledgements}
%\vspace{3mm}
\noindent
We are grateful to Gleb Arutyunov, Amit Dekel, Harald Dorn, Sergey Krivonos, Joaquim Gomis, Luka Megrelidze, Olof Ohlsson Sax, Bogdan Stefanski, Alessandro Torrielli and Kentaroh Yoshida for useful discussions. We also thank Olof Ohlsson Sax for comments on the manuscript.
M.H. thanks the organizers of the conference {\it Selected Topics in Theoretical High Energy Physics} (Tbilisi, 2015) 
and of the program {\it Holography and Dualities 2016: New Advances in String and Gauge Theory} (Nordita, 2016) as well as Nordita in Stockholm for kind hospitality.
G.J. thanks the Humboldt University of Berlin and University of Hamburg
for kind hospitality.
The work of M.H. and G.J. is supported  by the German Science Foundation (DFG) under the
Collaborative Research Center (SFB) 676 {\it Particles, Strings and the Early Universe}.
In addition, the work of G.J. is supported by the Rustaveli GNSF and the
DFG under the SFB 647 {\it Space--Time--Matter}.

\newpage

\appendix

\section{\texorpdfstring{More on $\SLS(2,\Reals)$}{SL2R} calculus}\label{app:SL2R}

	In this appendix we describe some technical details of $\SLS(2,\Reals)$ calculations.

	The basis vectors \eqref{sl(2) basis} satisfy the matrix relations 
	\be\label{tt=}
		\tt_\m \tt_\n=\eta_{\m\n}{\bf I}+\eps_{\m\n}\,^\rho \tt_\rho
	\ee
	and by \eqref{g=X1} one obtains
	\be\label{gt,tg}
		\tt_\n \,g=X^{0'}\tt_\n+X_\n{\bf I}-\eps_{\m\n}\,^\rho X^\m \tt_\rho~,
		\qquad
		g\,\tt_\n=X^{0'}\tt_\n+X_\n{\bf I}+\eps_{\m\n}\,^\rho X^\m \tt_\rho~.
	\ee
	From \eqref{g=X1} one also finds the infinitesimal transformations of $g$
	corresponding to \eqref{AdS isometry tr}
	\be\label{isometry tr of SL(2) 1}
		g\mapsto g+\a^{0'\n}(X^{0'}\tt_\n+X_\n{\bf I})+\a^{\rho\m}X_\m\tt_\rho~.
	\ee
	Introducing dual parameters $\b_\n$ by $\b_\n=-\f 1 2 \eps_{\n\rho\m}\a^{\rho\m}$,
	one gets $\a^{\rho\m}=\eps^{\rho\m\n}\,\b_{\n}$, due to $\eps^{\rho\m\n}\,\eps_{\n\rho'\m'}=\d^\rho_{\m'}\,\d^{\m}_{\rho'}- \d^\rho_{\rho'}\,\d^{\m}_{\m'}$, and
	\eqref{isometry tr of SL(2) 1} takes the form \eqref{SL(2) isometry tr}, because of \eqref{gt,tg}.

	Note that the infinitesimal conformal transformation \eqref{SL(2) conformal tr} has the form
	\be\label{SL(2) conformal tr 1}
		g\mapsto g+\veps^A K_A^\m \tt_\m\,g~,
	\ee
	which exhibits that the transformed matrix in \eqref{SL(2) conformal tr} is also an element of $\SLS(2,\Reals)$.
	Indeed, comparing \eqref{SL(2) conformal tr} and \eqref{SL(2) conformal tr 1} for the terms containing $\veps^{0'}$,
	we get $g^{-1}+X_{0'}{\bf I}=K_{0'}^\m \tt_\m$ and inserting here 
	$g^{-1}$ from \eqref{g=X1}, we find $K_{0'}^\m=-X^\m$. Similarly, the terms containing $\veps^\n$ yield 
	$\tt_\n\,g^{-1}+X_\n {\bf I}=K_\n^\m \tt_\m$ and lead to $K_\n^\m=X^{0'}\d^\m_n-\eps^\m\,_{\n\rho}X^\rho$.

	In practical calculations it is helpful to use the Chevalley basis of $\sls(2,\Reals)$ 
	\begin{equation}\label{sl(2) basis 2}
		{\bf{t}}_2=\begin{pmatrix} 1&0\\0&-1 \end{pmatrix},\qquad
			{\bf{t}}_+= \begin{pmatrix} 0&1\\0&0 \end{pmatrix},\qquad
			{\bf{t}}_-= \begin{pmatrix} 0&0\\1&0 \end{pmatrix},
	\end{equation}
	which satisfies the commutation relations 
	\be\label{sl2 algebra}
	[{\bf{t}}_2,{\bf{t}}_\pm]=\pm 2  {\bf{t}}_\pm~,\qquad\qquad [{\bf{t}}_+,{\bf{t}}_-]={\bf{t}}_2~,
	\ee
	and therefore $e^{\a\tt_2}\,\tt_\pm\,e^{-\a\tt_2}=e^{\pm2\a}\,\tt_+.$
	The nonzero components of the normalized Killing form are
	$\la \tt_2\,\tt_2 \ra=1,$  $\,\la \tt_+\,\tt_- \ra=\frac{1}{2}$.
	These equations simplify the calculation of the presymplectic forms \eqref{1-form m}, \eqref{1-form} and
	of the Noether charges \eqref{L,R in angles}, \eqref{L,R m=0 in angles}. 

	Using the parametrization \eqref{L=t+,R=t+}, the Noether charge related to the conformal transformations $C=g^{-1}\,L$ takes the following form  
	$C=g_r^{-1}\,\tt_+\,g_l^{-1}$. The Iwasawa decomposition \eqref{gl,gr} then leads to
	\be\label{C=}
		C=\begin{pmatrix}
			e^{\a_l}\,e^{-\a_r}\,\sin\th_l\,\cos\th_r & e^{\a_l}\,e^{-\a_r}\,\cos\th_l\,\cos\th_r\\
			e^{\a_l}\,e^{-\a_r}\,\sin\th_l\,\sin\th_r & e^{\a_l}\,e^{-\a_r}\,\cos\th_l\,\sin\th_r 
		\end{pmatrix},
	\ee
	and in the canonical coordinates \eqref{CC m=0} on obtains \eqref{C in pq}. This yields
	$\,C_{0'}=\f 1 2(q_l p_r+ p_l q_r)$, $\,C_{0}=\f 1 2(p_l p_r - q_l q_r)$,
	$\,C_{1}=-\f 1 2(q_l p_r+ q_l q_r)$, $\,C_{2}=\f 1 2(q_l p_r - p_l q_r)$.

	Now we prove that the Iwasawa decomposition \eqref{Iwasawa} for a given $g_l\in \SLS(2,\Reals)$ 
	uniquely fixes the parameters $\g_l\in \Reals^1$, $a_l\in \Reals^1$, and $\th_l\in S^1$. 
	For simplicity we omit the index $l$. First note that for a given  $g\in \SLS(2,\Reals)$ one can find the parameter $\g$ such that the matrix $\tilde g=e^{-\g\tt_+}\,g$ has rows orthogonal to each other. Indeed, for
	\be
		g=\begin{pmatrix} a & b\\ c & d \end{pmatrix} \qquad \text{one has}\qquad \tilde g=e^{-\g\tt_+}\,g=\begin{pmatrix} a-\g c & b-\g d\\ c & d~~~ \end{pmatrix}~,
	\ee
	and requiring orthogonality of the rows uniquely fixes the parameter $\g=\f{ac+bd}{c^2+d^2}$. Now, any such matrix $\tilde g$ can be parametrized as
	\be
		\tilde g = e^{\a\tt_2}\,e^{\th\tt_0}=\begin{pmatrix}
		e^\a\cos\th& e^\a\sin\th \\ -e^{-\a}\sin\th& e^{-\a}\cos\th \end{pmatrix}~.
	\ee
	Especially, as the two orthogonal rows are nonzero two-vectors, the parameters
	$\a$ and $\th$ are uniquely defined by one of the rows.
	The proof can easily be repeated for $g_r$ in  \eqref{Iwasawa}.

\section{Canonical map to the Holstein-Primakoff realization}
\label{app:HolPrim}

	Here we describe a canonical map which relates the Holstein-Primakoff type realization  of  
	$\sls(2,\Reals)$ to the realization given by \eqref{Hamiltonian}-\eqref{J_pm}. This map provides a one to one correspondence between complex canonical coordinates 
	$(b^*,b)$ on a plane and canonical coordinates $(p,q)$ given on the half-plane $q>0$.

	The Holstein-Primakoff realization classically  is represented as (see \eqref{Holstein-Pr})
	\be\label{sl2 generators 2}
	E= b^*b+m/2  , \quad  B^*=\sqrt{m+b^*b}\,b^* , \quad  B=\sqrt{m+b^*b}\,b~,
	\ee
	where $(b^*, b)$ are complex coordinates on $\Reals^2$ and
	the Poisson bracket $\{b,b^*\}=i$ provides the $\sls(2,\Reals)$ algebra
	\be\label{sl2 for E-B}
	\{ E, B^*\}=iB^*~, \qquad \{ E, B\}=-iB~, \qquad \{ B, B^*\}=2iE.
	\ee
	The generators \eqref{Hamiltonian}-\eqref{J_pm} classically can be written as
	\be\label{B1}
	H=\f 1 4\left(p^2+q^2 +{m^2}/{q^2}\right)~, \qquad 
	J^\pm=\f 1 4\left(p^2-q^2+{m^2}/{q^2}\right)\pm \f i 2\,\, pq
	\ee
	and the canonical bracket $\{p,q\}=1$ leads to the algebra
	\be\label{B2}
	\{H , J^\pm \}=\pm i\,J^\pm~, \qquad \{J^- , J^+ \}=2i\,H~,
	\ee
	which is equivalent to \eqref{sl2 for E-B} under the identifications 
	$E=H$, $\,B^*=-J^+$,  $\,B=-J^-$. 

	From these equations one finds 
	$q=\sqrt{2E+B^*+B},\,$  $\,pq=i(B^*-B)$ and the inverse map is given by
	\be\label{B3}
	b^*=-\f{J^+}{\sqrt {H+m/2}}~, \qquad b=-\f{J^-}{\sqrt {H+m/2}}~.
	\ee
	Then, from $\{p,q\}=1$ follows $\{b,b^*\}=i$ and vice versa, {\it i.e.} 
	$\dd p\wedge\dd q=i\,\dd b^*\wedge\dd b$.

\section{Parametrization of \texorpdfstring{$\OSP(1|2)$}{OSP12}}\label{app:OSP}

	In the last Appendix we describe parametrizations of $\OSP(1|2)$ group elements and give some
	useful formulas for calculations of the Noether charges and the symplectic forms.

	In the massive case one can start with $g_l=g_l^{(b)}\,g_l^{(f)}$, where $g_l^{(b)}$ and 
	$g_l^{(f)}$ are purely bosonic and purely fermionic parts, respectively. For the bosonic part we use the Iwasawa decomposition
	$g_l^{(b)}=e^{\g_l\,\TT_+}\,e^{\a_l\,\TT }\,e^{\th_l\,\TT_0}$,
	as in \eqref{Iwasawa}, and the fermionic part we represent in a symmetric form 
	$g_l^{(f)}=e^{\zeta_l\,{\bf S}_+ +\eta_l\,{\bf S}_-}$. Using then the relations
	\be\label{adj-tr}
		e^{\th\,\TT_0} \, {\bf S}_\pm\,e^{-\th\,\TT_0} =\cos\th\,{\bf S}_\pm \pm \sin\th\,{\bf S}_\mp~,
	\ee
	\be\label{C1}
		e^{\zeta\,{\bf S}_+ +\eta\,{\bf S}_-}=e^{\f 1 2 \eta\zeta\TT }
		e^{\zeta\,{\bf S}_+}\,e^{\eta\,{\bf S}_-}~,
	\ee
	which follow from the $\osp(1|2)$ algebra \eqref{osp(1.2) algebra},
	we represent $g_l$ as
	\be\label{gl for m>0}
	g_l=e^{\g_l\,\TT_+}\,e^{\tilde\a_l\,\TT }\,e^{\tilde\zeta_l\,{\bf S}_+}\,
	e^{\tilde\eta_l\,{\bf S}_-}\,e^{\th_l\,\TT_0}~,
	\ee
	where $\tilde{\zeta}_l=\cos\th_l \,\zeta_l-\sin\th_l\, \eta_l$,
	$\tilde{\eta}_l=\sin\th_l \,\zeta_l+\cos\th_l \, \eta_l$ and $\tilde\a_l=\a_l+
	\f 1 2\,\tilde\eta_l\,\tilde\zeta_l$. Removing then `tilde' in \eqref{gl for m>0}, we obtain $g_l$ in \eqref{SUSY g_l-g_r m}. 
	$g_r$  is obtained in a similar
	way, starting with $g_r=g_r^{(f)}\,g_r^{(b)}$ and using the same steps as for $g_l$.

	The calculation of the Noether charges and the presymplectic forms in the massive case 
	(see \eqref{SUSY L in angles}-\eqref{1-form l})
	is based on the following relations 
	\be\ba\label{C2}
		&e^{\zeta\,{\bf S}_+}\,\TT_0\,e^{-\zeta\,{\bf S}_+}=\TT_0-\zeta\,{\bf S}_-~,& \quad
		&e^{\eta\,{\bf S}_-}\,\TT_0\,e^{-\eta\,{\bf S}_-}=\TT_0+\eta\,{\bf S}_+~,\\
		&e^{\zeta\,{\bf S}_+}\,\eta\,{\bf S}_+\,e^{-\zeta\,{\bf S}_+}=\eta\,{\bf S}_+ + 2\,\zeta\eta\,\TT_+~,&   &e^{\g\,\TT_+}\,\TT_-\,e^{-\g\,\TT_+}=
		\TT_-+\g\,\TT -\g^2\TT_+~,
	\ea\ee
	which also follow from \eqref{osp(1.2) algebra}.
	
	In the massless case we start again with $g_l=g_l^{(b)}\,g_l^{(f)}$,
	where $g_l^{(b)}=e^{\th_l\,\TT_0}\,e^{\a_l\,\TT }\,e^{\g_l\,\TT_+}$ and
	$g_l^{(f)}=e^{\zeta_l\,{\bf S}_- +\eta_l\,{\bf S}_+}$.
	Similarly to \eqref{adj-tr}-\eqref{C1}, here we use
	\be\label{adj-tr 1}
		e^{\g\,\TT_+} \, {\bf S}_+\,e^{-\g\,\TT_+} ={\bf S}_+~, \qquad
		e^{\g\,\TT_+} \, {\bf S}_-\,e^{-\g\,\TT_+} ={\bf S}_--\g\,{\bf S}_+~,
		\ee
		\be\label{C11}
		e^{\zeta\,{\bf S}_- +\eta\,{\bf S}_+}=e^{\f 1 2 \eta\zeta\TT }
		e^{\zeta\,{\bf S}_-}\,e^{\eta\,{\bf S}_+}~,
	\ee
	which leads to
	\be\label{C3}
		g_l=e^{\th_l\,\TT_0}\,e^{\tilde\a_l\,\TT }\,e^{\zeta_l\,{\bf S}_-}\,e^{\tilde\eta_l\,{\bf S}_+} \,e^{\g_l\,\TT_+}~, 
	\ee
	where $\tilde\eta_l=\eta_l-\g_l\,\zeta_l$ and $\tilde{\a}_l=\a_l+\f 1 2\,\eta_l\zeta_l$.
	Removing again the 'tilde', we get $g_l$ in \eqref{SUSY gl,gr}. The parametrization 
	of $g_r$ is derived in a similar way.

	Finally, we present some
	helpful formulas for the calculations in the massless case
	\be\ba\label{C4}
	&e^{\eta\,{\bf S}_+}\,\TT_+\,e^{-\eta\,{\bf S}_+}=\TT_+~,& \quad
	&e^{\zeta\,{\bf S}_-}\,\TT_+\,e^{-\zeta\,{\bf S}_-}=\TT_+ +\zeta\,{\bf S}_+~,\\
	&e^{\a\,\TT }\,\TT_\pm\,e^{-\a\,\TT }=e^{\pm 2\a}\,\TT_\pm 
	~,&   &e^{\a\,\TT }\,{\bf S}_\pm\,e^{-\a\,\TT }=e^{\pm \a}\,{\bf S}_\pm~,
	\ea\ee
	\be\label{adj-tr 3}
	e^{\th\,\TT_0} \, \TT_+\,e^{-\th\,\TT_0} =\cos^2\th\,\TT_+ 
	-\sin^2\th\,\TT_- +\sin\th\,\cos\th \,\TT ~.
	\ee

%%%%%%%%%%%%%%%%%%%%%%%%%%%%%%%%%%%%%%%%%%%%%%%%%%%%%%%%%%%%%%%%%%%%%%%%%%%%%%%%
%%%%%%%%%%%%%%%%%%%%%%%%%%%%%%%%%%%%%%%%%%%%%%%%%%%%%%%%%%%%%%%%%%%%%%%%%%%%%%%%
\bibliographystyle{../../../../Latex/bibSpires/nb}
\bibliography{../../../../Latex/bibSpires/bibSpires}
%\bibliographystyle{nb}
%\bibliography{bibSpires}

\begin{comment}

\end{comment}
  
\end{document}